\documentclass[12pt]{article}
\usepackage[utf8]{inputenc}
\usepackage{verbatim,graphics,graphicx,color,slashed,amsmath}
\usepackage[T1]{fontenc}
\usepackage{subfigure}
\usepackage[table,svgnames]{xcolor}
\usepackage{xcolor}
\usepackage{hyperref}
\usepackage{amsfonts}
\usepackage{slashed}
\usepackage{amssymb}
\usepackage{bbding}
\usepackage{appendix}
\usepackage{mhchem}
\usepackage{stmaryrd}
\usepackage{blkarray}
\usepackage[export]{adjustbox}
\graphicspath{ {./images/} }
\usepackage{bbold}
\usepackage{CJKutf8}
\usepackage[T1]{fontenc}
\usepackage{cancel}
\usepackage{enumerate}
\usepackage{cite}
\allowdisplaybreaks[1]

\def\thefootnote{\fnsymbol{footnote}}

\usepackage{multirow}

\newcommand{\be}{\begin{equation}}
\newcommand{\ee}{\end{equation}}
\newcommand{\bea}{\begin{eqnarray}}
\newcommand{\eea}{\end{eqnarray}}

\usepackage[normalem]{ulem}

\newcommand{\mc}{\mathcal}
\newcommand{\tr}{\text{tr}}

\RequirePackage[normalem]{ulem} 
\RequirePackage{color}\definecolor{RED}{rgb}{1,0,0}\definecolor{BLUE}{rgb}{0,0,1} 

\begin{document}

\begin{center}
{\Large\bf One-loop Matching and Running via On-shell Amplitudes}
\end{center}

\vspace{0.2cm}

\begin{center}
{\bf Xu Li}~\footnote{E-mail: lixu96@ihep.ac.cn},
\quad
{\bf Shun Zhou}~\footnote{E-mail: zhoush@ihep.ac.cn}
\\
\vspace{0.2cm}
{\small
Institute of High Energy Physics, Chinese Academy of Sciences, Beijing 100049, China\\
School of Physical Sciences, University of Chinese Academy of Sciences, Beijing 100049, China}
\end{center}

\vspace{1.5cm}

\begin{abstract}
In this work, we put forward a straightforward and simple approach to construct the low-energy effective field theory (EFT) from a given ultraviolet (UV) full theory by integrating heavy particles out. By calculating the on-shell amplitudes, we demonstrate how to directly achieve the one-loop matching of the UV full theory onto the EFT with the complete set of independent operators in the physical basis, which are usually obtained by removing the redundant operators in the Green's basis via the equations of motion. Furthermore, taking specific examples, we explain how to implement the on-shell-amplitude approach to derive the one-loop renormalization-group equations for the Wilson coefficients, and to find out the contributions from the evanescent operators.
\end{abstract}


\def\thefootnote{\arabic{footnote}}
\setcounter{footnote}{0}
\newpage

\section{Introduction}
\label{sec1}
The effective field theories (EFTs) have been extensively studied and widely applied in condensed matter physics, particle physics, nuclear physics, gravity and cosmology. In particle physics, the standard model effective field theory (SMEFT) serves as a powerful tool to explore new physics beyond the standard model (SM) in a model-independent way, as the impact of new physics on the low-energy phenomena of elementary particles has been contained in all the higher-dimensional operators and the associated Wilson coefficients~\cite{Buchmuller:1985jz,Isidori:2023pyp}. In fact, the SMEFT has been utilized to analyze the experimental data, and restrictive constraints on the Wilson coefficients and the energy scale of possible new physics have been extracted~\cite{Ellis:2018gqa,Ethier:2021bye,deBlas:2022ofj}.
On the other hand, motivated by the unresolved problems in the SM, such as neutrino masses, the stability of the electroweak scale and the existence of dark matter, concrete new-physics models have been proposed with various new particles that may be too heavy to be directly produced in collider experiments. In this case, to confront these ultraviolet (UV)-complete models with low-energy precision data, one may integrate out heavy particles in the UV full theory and match it onto the SMEFT. Once deriving the higher-dimensional operators and their Wilson coefficients, one should further calculate the beta functions of relevant coefficients and determine their values at low energies by solving the renormalization-group (RG) equations~\cite{Babu:1993qv,Chankowski:1993tx,Antusch:2001ck,Broncano:2004tz,Jenkins:2013zja,Jenkins:2013wua,Alonso:2013hga,Alonso:2014zka,Liao:2016hru,Liao:2019tep,Chala:2021juk,Chala:2021pll,DasBakshi:2022mwk,DasBakshi:2023htx}.

Since the SMEFT has been adopted as the theoretical framework to investigate the deviations from the SM predictions for low-energy observables, a complete set of independent operators up to a given mass dimension is obviously necessary~\cite{Weinberg:1979sa,Grzadkowski:2010es,Lehman:2014jma,Li:2020gnx,Murphy:2020rsh,Liao:2020jmn,Li:2020xlh,Harlander:2023psl}. At the dimension-six (dim-6) level, such a complete set of operators is known as the Warsaw basis~\cite{Grzadkowski:2010es}, which will henceforth be referred to as the ``physical basis'' in the sense of no redundant operators. Given any UV-complete theory or EFT with particles heavier than the energy scale of interest, one can further integrate out heavy particles and construct the EFT at next level. In the ordinary way, one first obtains all possible operators allowed by symmetries and then use the Dirac algebra, Fierz identities, and integration by parts to convert these operators into those in the Green's basis~\cite{Jiang:2018pbd,Gherardi:2020det,Chala:2021cgt, Ren:2022tvi,Zhang:2023kvw}.
Compared to the operators in the physical basis, those in the Green's basis are redundant up to the equations of motion (EOMs) of fields. For clarity, we shall simply call such redundance in the Green's basis as the ``EOM redundance.''

In the literature, the one-loop matching~\cite{Jiang:2018pbd,Gherardi:2020det,Chala:2020vqp,Haisch:2020ahr,Dittmaier:2021fls,Zhang:2021tsq,Zhang:2021jdf,Coy:2021hyr,Ohlsson:2022hfl,Li:2022ipc,Du:2022vso,Zhang:2022osj,Liao:2022cwh,Fuentes-Martin:2020udw,Carmona:2021xtq,Fuentes-Martin:2022jrf} and running of Wilson coefficients~\cite{Jenkins:2013zja,Jenkins:2013wua,Alonso:2013hga,Alonso:2014zka,Liao:2016hru,Liao:2019tep,Chala:2021juk,Chala:2021pll,DasBakshi:2022mwk,DasBakshi:2023htx} have usually been performed by first calculating the off-shell amplitudes in the Green's basis and then manually removing the EOM redundance. Although the operators in the physical basis of the SMEFT have been provided up to mass dimension twelve, one still needs to construct the Green's basis in practice. An immediate question is whether one can directly carry out the one-loop matching and calculate the beta functions in a single step without explicitly dealing with the EOMs. In this work, we put forward a straightforward and simple approach to do so. The essential idea is to calculate the on-shell amplitudes for any given number of external light fields, and then match them with those contributed by the operators in the physical basis. In this way, only the physical basis is needed, while the Green's basis and the EOM redundance are irrelevant. A similar strategy has also been mentioned briefly in Sec. 2.4 of Ref.~\cite{Aebischer:2023irs}.

The remaining part of this paper is structured as follows. In Sec.~\ref{sec:convention}, we first recall the standard procedure of the one-loop off-shell matching, and then present our new method to achieve this with on-shell amplitudes. A proof is given in Sec.~\ref{sec:proof} to validate the on-shell-amplitude approach. In Sec.~\ref{sec:application}, this approach is then applied to the one-loop matching of a toy UV model onto its corresponding EFT, and the matching conditions are derived. In addition, we re-derive the beta functions of the Wilson coefficients of several dim-6 operators in the Warsaw basis, and treat the contributions from the evanescent operators to the one-loop matching in the seesaw effective field theories. Finally, we summarize our main results and conclude in Sec.~\ref{sec:conclude}.

\section{The On-shell-amplitude Approach}
\label{sec:convention}

In this section, we first introduce the one-loop off-shell matching in the path-integral formalism, and then present the general idea of the on-shell-amplitude approach. To begin with, we explain our notations and conventions that will be used in subsequent discussions. For a general field theory, the generating functional for the Green's functions is given by
\bea
Z[J^{}_\phi]=\int{\mc{D}\phi} \exp\left\{ {\rm i} \int{{\rm d}^4 x} \left( \mc{L}[\phi]+J^{}_\phi \phi \right) \right\} \; ,
\eea
where ${\cal L}[\phi]$ denotes the Lagrangian and $J^{}_\phi$ stands for the external source coupled to the field $\phi$. Note that the spacetime coordinates of $J^{}_\phi(x)$ and $\phi(x)$ have been suppressed. The generating functional for the connected Green's functions is $W[J^{}_\phi] \equiv -{\rm i} \ln Z[J^{}_\phi]$, through which one can define the mean field $\overline{\phi} \equiv \delta W[J^{}_\phi]/\delta J^{}_\phi$ in the presence of the external source $J^{}_\phi$. Then, the effective action $\Gamma[\overline{\phi}] \equiv W[J^{}_\phi] - \int {\rm d}^4x J^{}_\phi \overline{\phi}$ is defined via the Legendre transformation and generates the one-particle irreducible (1PI) $n$-point vertex, i.e.,
\bea
\Gamma^{\left(n\right)}\left(x^{}_1,\cdots,x^{}_n\right) \equiv \frac{{\rm i} \delta^n \Gamma\left[\overline{\phi}\right]}{\delta\overline{\phi}^{}_{x^{}_1}\cdots \delta \overline{\phi}^{}_{x^{}_n}} \;,
\eea
where $\overline{\phi}^{}_{x^{}_i} \equiv \overline{\phi}(x^{}_i)$ has been defined for $i = 1, 2, \cdots, n$. The representation of $\Gamma^{\left(n\right)}$ in the four-momentum space can be obtained from the Fourier transform
\bea
\Gamma^{\left(n\right)}\left(x_1,\cdots,x_n\right)=\int{\frac{{\rm d}^4k^{}_1}{\left(2\pi\right)^4}\cdots\frac{{\rm d}^4k^{}_n}{\left(2\pi\right)^4}}e^{{\rm i} \left(k^{}_1 x^{}_1 + \cdots + k^{}_n x^{}_n\right)} \Gamma^{\left(n\right)}\left(k^{}_1, \cdots, k^{}_n \right) \left(2\pi\right)^4 \delta^{4} \left(k^{}_1 + \cdots + k^{}_n\right) \;. ~~
\eea
In this way, the effective action can be expanded in terms of the $n$-point vertices either in the coordinate space or in the momentum space as
\bea
{\rm i}\Gamma\left[\overline{\phi}\right] &=& \sum_{n}\frac{1}{n!}\int{{\rm d}^4 x^{}_1\cdots {\rm d}^4 x^{}_n}\ \Gamma^{\left(n\right)}\left(x^{}_1, \cdots, x^{}_n\right) \overline{\phi}^{}_{x^{}_1} \cdots \overline{\phi}^{}_{x^{}_n} \nonumber \\
&=& \sum_{n} \frac{1}{n!} \int{\frac{{\rm d}^4 k^{}_1}{\left(2\pi\right)^4} \cdots \frac{{\rm d}^4 k^{}_n}{\left(2\pi\right)^4}}\ \left(2\pi\right)^4\delta^{4} \left(k^{}_1 + \cdots + k^{}_n\right) \Gamma^{\left(n\right)}\left(k^{}_1, \cdots, k^{}_n\right) \overline{\phi}^{}_{k^{}_1} \cdots \overline{\phi}^{}_{k^{}_n} \;,
\label{eq:Gamma}
\eea
where the field $\overline{\phi}^{}_k \equiv \overline{\phi}(k)$ in the momentum space is related to that in the coordinate space via the Fourier transform $\overline{\phi}^{}_k \equiv \int{{\rm d}^4 x} \, e^{{\rm i}k x} \overline{\phi}^{}_x$. In the following discussions, we will mainly focus on the effective vertices in the momentum space.

To illustrate the standard procedure for one-loop matching of the UV model onto the EFT, we consider a toy model that contains a light real scalar $\phi$ and a heavy one $\Phi$, for which the generating functional is given by
\bea
Z\left[J^{}_\phi, J^{}_\Phi\right] = \int \mc{D}\phi \mc{D}\Phi \exp\left\{{\rm i} \int{\rm d}^4 x \left(\mc{L}^{}_{\rm UV}\left[\phi, \Phi\right] + J^{}_\phi \phi + J^{}_\Phi \Phi\right)\right\} \;.
\eea
Using the background-field method, one can divide these scalar fields into the background fields and quantum fluctuation fields as $\phi \rightarrow \phi^{}_{\rm B} +\phi^\prime$ and $\Phi \rightarrow \Phi^{}_{\rm B} + \Phi^\prime$, so that the path integral is performed only over the quantum fields. To this end, the Lagrangian can be expanded around the background fields up to the second order of quantum fields as
\bea
\mc{L}^{}_{\rm UV}\left[\phi, \Phi\right] + J^{}_\phi \phi + J^{}_\Phi \Phi
\simeq \mc{L}^{}_{\rm UV} \left[\phi^{}_{\rm B}, \Phi^{}_{\rm B} \right] + J^{}_\phi \phi^{}_{\rm B} + J^{}_\Phi \Phi^{}_{\rm B} - \frac{1}{2} \left(\Phi', \phi'\right)\mc{Q}^{}_{\rm UV}\left(\begin{matrix}\Phi'\\\phi'\\\end{matrix}\right) \;,
\label{eq:Lexpand}
\eea
where the first-derivative terms disappear due to the EOMs of background fields, and $\mc{Q}^{}_{\rm UV}$ collects the second-derivative terms, i.e.,
\bea
\mc{Q}^{}_{\rm UV} \equiv \left(\begin{array}{cc}
	\displaystyle -\frac{\delta^2 \mathcal{L}_{\mathrm{UV}}}{\delta \Phi^2}\left[\Phi_{\mathrm{B}}, \phi_{\mathrm{B}}\right] & \displaystyle -\frac{\delta^2 \mathcal{L}_{\mathrm{UV}}}{\delta \Phi \delta \phi}\left[\Phi_{\mathrm{B}}, \phi_{\mathrm{B}}\right] \\
	\displaystyle -\frac{\delta^2 \mathcal{L}_{\mathrm{UV}}}{\delta \phi \delta \Phi}\left[\Phi_{\mathrm{B}}, \phi_{\mathrm{B}}\right] & \displaystyle -\frac{\delta^2 \mathcal{L}_{\mathrm{UV}}}{\delta \phi^2}\left[\Phi_{\mathrm{B}}, \phi_{\mathrm{B}}\right]
	\end{array}\right) \;.
	\label{eq:QUV}
\eea
Notice that the background field $\phi^{}_{\rm B}$ (or $\Phi^{}_{\rm B}$) coincides with the mean field $\overline{\phi}$ (or $\overline{\Phi}$) at the classical level, and quantum corrections to background fields result in the changes at the two-loop order in the effective action. Now we integrate $\Phi'$ out in the UV theory, and define the one-light-particle irreducible (1LPI) effective action $\Gamma_{\rm L,UV}[\phi]$ as the Legendre transformation of the generating functional of the connected Green's functions with $J^{}_\Phi=0$. For $J^{}_\Phi=0$, the classical field ${\Phi}^{}_{\rm c}[\phi]$ satisfies the EOM, i.e., $(\delta \mc{L}/\delta\Phi)|_{J^{}_\Phi=0} = 0$. Since ${\Phi}^{}_{\rm c}[\phi]$ is usually a non-local functional of $\phi$, it can be expanded in terms of $1/M^{}_\Phi$ up to the desired power to obtain the local one $\hat{\Phi}[\phi]$, where $M^{}_\Phi$ is the heavy-particle mass. Eventually, the 1LPI effective action $\Gamma^{}_{\rm L,UV}[\phi]$ with the local field $\hat{\Phi}[\phi]$ substituted becomes
\bea
\Gamma^{}_{\rm L,UV}\left[\phi^{}_{\rm B}\right]&=&-{\rm i}\ln{Z\left[J^{}_\Phi=0, J^{}_\phi\right]}-\int {\rm d}^4x J^{}_\phi \phi^{}_{\rm B} \\
&= &\int{{\rm d}^4x} \mc{L}^{}_{\rm UV} [\phi^{}_{\rm B}, \hat{\Phi}[\phi^{}_{\rm B}]] + \frac{\rm i}{2}\ln{\det{\mc{Q}^{}_{\rm UV}[\phi^{}_{\rm B},\hat{\Phi}[\phi^{}_{\rm B}]]}} \;,
\eea
where the last term is obtained by the Gaussian integration of $\phi'$ and $\Phi'$ over the last term in Eq.~\eqref{eq:Lexpand}. As a consequence, the leading- and one-loop-order parts of the 1LPI effective action are respectively given by
\bea
\Gamma_{\rm L,UV}^{(0)}\left[\phi^{}_{\rm B}\right] &=& \int{{\rm d}^4x}\mc{L}\left[\phi^{}_{\rm B},\hat{\Phi}[\phi^{}_{\rm B}]\right] \;, \\
\Gamma_{\rm L,UV}^{(1)}\left[\phi^{}_{\rm B}\right] &=& \frac{\rm i}{2} \ln \det \mc{Q}^{}_{\rm UV} \left[\phi^{}_{\rm B}, \hat{\Phi}[\phi^{}_{\rm B}]\right] \;.
\eea
In the EFT, there is only the light field $\phi$, so the effective action can be simply written as
\bea
\Gamma^{}_{\rm EFT}\left[\phi^{}_{\rm B}\right] = \int {\rm d}^4x \left( \mc{L}_{\rm EFT}^{(0)} \left[\phi^{}_{\rm B}\right] + \mc{L}_{\rm EFT}^{(1)}\left[\phi^{}_{\rm B}\right] \right) + \frac{\rm i}{2} \ln{\det{\mc{Q}_{\rm EFT}}} \;.
\eea
The construction of $\mc{Q}^{}_{\rm EFT}$ is similar to that in Eq.~\eqref{eq:QUV}, namely, $\mc{Q}^{}_{\rm EFT} = -\delta^2 \mc{L}_{\rm EFT}^{(0)}/\delta\phi^2$. The one-loop matching is realized by requiring that the 1LPI effective action in the UV full theory to be equal to the 1PI effective action in the EFT at the energy scale characterized by the heavy-particle mass, i.e., $\Gamma^{}_{\rm L,UV}\left[\phi^{}_{\rm B}\right] = \Gamma^{}_{\rm EFT}\left[\phi^{}_{\rm B}\right]$. Furthermore, it can be proved~\cite{Zhang:2016pja} that $\ln{\det{\mc{Q}^{}_{\rm EFT}}} = \ln{\det{\mc{Q}^{}_{\rm UV}[\phi^{}_{\rm B},\hat{\Phi}^{} [\phi_{\rm B}]]}}|_{\rm soft}$, where the subscript ``soft'' indicates the soft region of the loop momentum according to the method of expansion by regions~\cite{Beneke:1997zp,Jantzen:2011nz}.
Therefore, only the hard-region part in $\Gamma_{\rm L,UV}^{(1)}$ contributes to the EFT effective action and the one-loop matching conditions are
\bea
\Gamma_{\rm L,UV}^{(0)}\left[\phi\right]&=& \int{{\rm d}^4 x}\mc{L}_{\rm EFT}^{(0)}\left[\phi\right]\bigg|_{\rm Green} \;, \nonumber \\
\Gamma_{\rm L,UV}^{(1)}\left[\phi\right]\big|_{\rm hard} &=& \int{{\rm d}^4 x}\mc{L}_{\rm EFT}^{(1)}\left[\phi\right]\bigg|_{\rm Green} \;,
\label{eq:matching}
\eea
where the operators in the EFT on the right-hand side (r.h.s.) are given in the Green's basis and we henceforth drop the subscript of the light background field $\phi^{}_{\rm B}$. Notice that the conditions in Eq.~(\ref{eq:matching}) apply in general to the theories with scalar bosons, vector bosons and fermions~\cite{Zhang:2016pja}, though we just take the scalar field $\phi$ for illustration.

Since the effective action is the generating functional for the 1LPI Green's functions, we can apply it to any specific off-shell processes with $n$ external legs, which can be expressed in the momentum space as
\bea
\frac{{\rm i}\delta^{n} \Gamma^{}_{\rm L,UV}\left[\phi\right]}{\delta\phi^{}_{p^{}_1} \ldots \delta\phi^{}_{p^{}_n}} \bigg|_{\rm hard} = \frac{{\rm i}\delta^{n}  }{\delta\phi^{}_{p^{}_1} \ldots \delta\phi^{}_{p^{}_n}} \int{{\rm d}^4 x} \left(\mc{L}_{\rm EFT}^{(0)}\left[\phi\right] + \mc{L}_{\rm EFT}^{(1)}\left[\phi\right] \right) \bigg|_{\rm Green} \;.
\label{eq:nleg}
\eea
It is worthwhile to mention that the direction of all momenta in Eq.~(\ref{eq:nleg}) is assumed to be incoming, and the subscript ``$|_{\rm hard}$'' on the left-hand side (l.h.s.) refers to the hard-region part of $\Gamma_{\rm L,UV}^{(1)}\left[\phi\right]$. As shown in Eq.~(\ref{eq:matching}), the off-shell 1LPI amplitudes in the UV theory are matched with the amplitudes generated by the operators in the EFT in the Green's basis. The Green's basis for the operators composed of $\phi$ may contain redundant operators that always appear as $F[\phi](D^2\phi)^m$ with $m$ being any positive integer and $D^2 \equiv D^{}_\mu D^\mu$ being square of the covariant derivative, where $F[\phi]$ is a general functional of $\phi$ but not its derivative $D^2\phi$. In the previous works, to convert the Green's basis to the physical basis, one has to resort to the EOM or the field redefinition of $\phi$ and remove all the redundant operators $F[\phi](D^2\phi)^m$ up to a given mass dimension.
\begin{figure}
	\bea
	\begin{gathered}
		\includegraphics[width=0.18\linewidth]{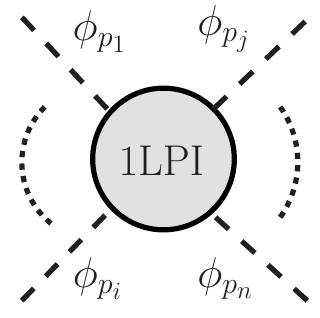}
	\end{gathered}
	\  + \
	\begin{gathered}
		\includegraphics[width=0.34\linewidth]{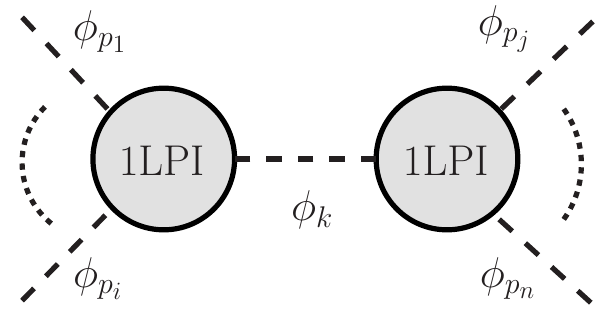}
	\end{gathered} \ + \cdots  + \
	\begin{gathered}
		\includegraphics[width=0.275\linewidth]{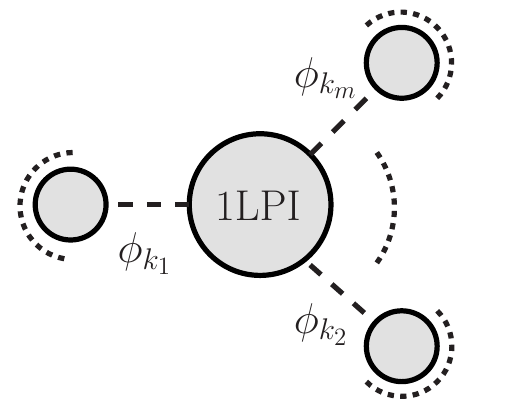}
	\end{gathered}
	\nonumber
	\eea
	\caption{The Feynman diagrams need to be considered in the on-shell-amplitude approach. The first diagram corresponds to Eq.~\eqref{eq:nleg} but with the on-shell conditions imposed, and the second one accounts for the redundant term $F[\phi] D^2\phi$ while the last one for the terms $F[\phi](D^2\phi)^m$.}
	\label{maindiag}
\end{figure}

One immediate question is whether one can derive all the independent operators in the physical basis together with their Wilson coefficients in the EFT directly from the UV theory. In this work, we provide an affirmative answer and a straightforward approach to do so. Before justifying the validity of our approach, we explain two key points to achieve this goal.
\begin{itemize}
	\item First, we impose two on-shell conditions on Eq.~\eqref{eq:nleg} to obtain the physical amplitude~\cite{Cohen:2022uuw}. In the momentum space, these two conditions are
	\bea
	\frac{{\rm i}\delta\Gamma}{\delta\phi^{}_p} &=& J^{}_p = 0\;, \label{eq:eom} \\
	\frac{{\rm i} \delta^2\Gamma}{\delta\phi^{}_{p^{}_i} \delta\phi^{}_{p^{}_j}} \bigg|_{J^{}_p=0} &=& -\left(2\pi\right)^4\delta^{4}\left(p^{}_i+p^{}_j\right)D^{-1}\left(p^{}_i\right) \big|_{J^{}_p=0} = 0 \;,	\label{eq:onshell}
	\eea
	where $J^{}_p$ is the source of $\phi^{}_p$ in the momentum space and ${\rm i}D^{-1}(p)=p^2-m^2_0 + \hat{\Sigma}(p^2)$ with $m^{}_0$ being the bare mass and $\hat{\Sigma}(p^2)$ being the 1PI self-energy correction. It is worth noting that Eq.~(\ref{eq:eom}) refers to the EOM of $\phi^{}_p$, while Eq.~(\ref{eq:onshell}) is the on-shell condition for the external leg. At $J_p=0$, Eq.~\eqref{eq:onshell} implies $\hat{\Sigma}(\hat{m}^2) = m^2_0 - \hat{m}^2$, where $\hat{m}$ denotes the pole mass of the light particle.
	
	\item Second, as the field redefinition of $\phi$ in the EFT can transform the operators on the r.h.s of Eq.~\eqref{eq:nleg} in the Green's basis to those in the physical basis, we just apply the same redefinition on the l.h.s of Eq.~\eqref{eq:nleg}. As a consequence, not only the 1LPI diagrams in the UV theory are relevant, but also reducible diagrams need to be taken into account. More explicitly, we have to consider all the diagrams in Fig.~\ref{maindiag} and calculate their on-shell amplitudes. The first diagram in Fig.~\ref{maindiag} corresponds to Eq.~\eqref{eq:nleg} but with the on-shell conditions imposed. The second diagram is needed to include the contribution of converting the redundant operators $F[\phi] D^2\phi$ to those in the physical basis. The last diagram is similar to the second one but now for the terms $F[\phi] (D^2\phi)^m$. The reducible amplitudes may contain the terms where the square of the four-momentum $k_i^2$ appears in the denominator. These terms mean non-locality, and thus should be dropped when matching the amplitudes to those in the EFT, in which all the operators are local.
\end{itemize}

\begin{figure}[t]
	\centering
	\includegraphics[width=0.55\linewidth]{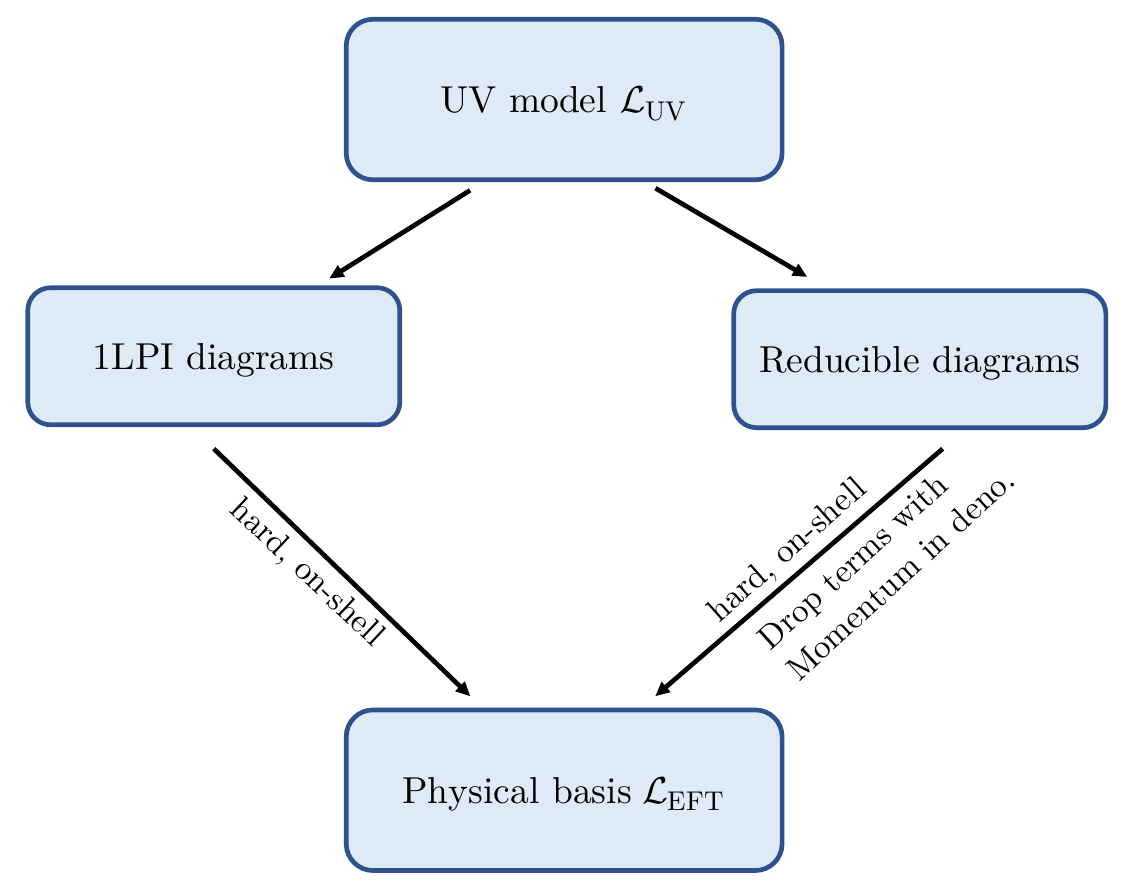}
	\caption{The flow chart of the generic procedure for the one-loop matching by using the on-shell-amplitude approach.}
	\label{procedure}
\end{figure}

The flow chart of the working procedure is displayed in Fig.~\ref{procedure}, where we explain how to start from the UV full theory to derive the Wilson coefficients of the independent operators in the physical basis of the EFT. Three basic steps are summarized below.
\begin{enumerate}
	\item Draw all possible Feynman diagrams and calculate the corresponding amplitudes. For the one-loop matching of the UV theory onto the EFT, it only needs to extract the hard-momentum contributions.
	\item Drop all the amplitudes that contain external momenta in the denominator, and then employ the on-shell conditions to obtain the final amplitude ${\cal M}^{\prime}_{\rm UV}$ in the UV theory. For example, for the $2$-to-$2$ process, a term like $t/s$ needs to be directly discarded, where $s$ and $t$ are the Mandelstam variables.
	\item Calculate the tree-level amplitude ${\cal M}_{\rm EFT}^{(0)}$ in the physical basis of the EFT, and identify it with the previously obtained amplitude in the UV theory, namely, ${\cal M}^{\prime}_{\rm UV}= {\cal M}_{\rm EFT}^{(0)}$. The Wilson coefficients of the operators in the EFT are thus determined.
\end{enumerate}
Given the complete and independent operators in the physical basis of the EFT, such as the Warsaw basis of dim-6 operators for the SMEFT, one can directly perform the one-loop matching of any UV full theory onto the SMEFT. Furthermore, such a procedure can also be applied to the EFT instead of the UV theory and calculate the beta functions of the Wilson coefficients.

\section{Proof and Generalization}
\label{sec:proof}

Now we shall prove that the on-shell-amplitude approach and the working procedure presented in the previous section are indeed correct. In this section, we give the proof in two steps. First, we briefly review the relationship between the implementation of the EOM and that of the field redefinition, both of which can be used to remove the EOM redundance in the EFT. Second, after fixing the exact form of the field redefinition in the EFT, we apply the same field redefinition in the UV theory, which will result in more reducible diagrams that need to be considered in addition to the 1LPI ones. Finally, apart from scalar fields, the proof will be generalized to the UV theories with fermion and gauge fields.

\subsection{EOM and field redefinition}
For simplicity, we consider the EFT with a massless real scalar $\phi$ and assume that only one redundant operator $F[\phi]\partial^2\phi$ is present. The relevant Lagrangian is written as
\bea
\mc{L}^{}_{\rm EFT}\left[\phi\right] &=& \frac{1}{2}\partial^{}_\mu\phi \partial^\mu\phi + V\left[\phi\right] + \frac{F\left[\phi\right]}{\Lambda^2}\partial^2\phi\;,
\label{eq:redunL}
\eea
where $F[\phi]$ is self-conjugate and assumed to be of dim-3, and $\Lambda$ is the cutoff scale. Note that both $V[\phi]$ and $F[\phi]$ are general and non-redundant, where $V[\phi]$ may include higher-dimensional operators. It is well known that the last term in Eq.~\eqref{eq:redunL} can be removed with the help of the EOM, namely, $\partial^2\phi = {\rm d} V/{\rm d}\phi$.

Instead of using the EOM, we make the field redefinition $\phi \rightarrow \phi + F[\phi]/\Lambda^2$ and substitute it back into Eq.~\eqref{eq:redunL}. Then, the Lagrangian can be recast into
\bea
\mc{L}^{}_{\rm EFT}\left[\phi\right]&=&\frac{1}{2}\partial^{}_\mu\phi \partial^\mu\phi+V\left[\phi+\frac{F[\phi]}{\Lambda^2}\right]+\mc{O}\left(\frac{1}{\Lambda^4}\right) \\
&\approx& \frac{1}{2}\partial^{}_\mu\phi\partial^\mu\phi +V\left[\phi\right]+\frac{{\rm d}V}{{\rm d}\phi}\frac{F[\phi] }{\Lambda^2}+\mc{O}\left(\frac{1}{\Lambda^4}\right)\;. \label{eq:redef}
\eea
Comparing Eq.~(\ref{eq:redef}) with Eq.~\eqref{eq:redunL}, one can observe that the field redefinition is actually equivalent to using the EOM, i.e.,  $\partial^2\phi \rightarrow {\rm d}V/{\rm d}\phi$, in the original Lagrangian in Eq.~\eqref{eq:redunL}.

\subsection{Field redefinition in the UV functional}
Now that the field redefinition $\phi \to \phi + F[\phi]/\Lambda^2$, which has been intended for removing the EOM redundance of the r.h.s of Eq.~\eqref{eq:nleg}, is determined, we apply the same field redefinition on the l.h.s of Eq.~\eqref{eq:nleg}. In this way, the effective action can be simply expanded as
\bea
\Gamma^{}_{\rm L,UV} \left[\phi + \frac{F[\phi]}{\Lambda^2}\right] &=& \Gamma^{}_{\rm L,UV}\left[\phi\right] + \int{{\rm d}^4 x\left(\frac{\delta\Gamma^{}_{\rm L,UV}\left[\phi\right]}{\delta\phi^{}_x} \cdot \frac{F[\phi^{}_x]}{\Lambda^2}\right)} +\mc{O}\left(\frac{1}{\Lambda^4}\right) \nonumber \\
&\approx &\Gamma^{}_{\rm L,UV}\left[\phi\right] + \int{{\rm d}^4 x} \int{\frac{{\rm d}^4 k}{(2\pi)^4}}\left(\frac{\delta\Gamma^{}_{\rm L,UV}\left[\phi\right]}{\delta\phi^{}_{-k}} \cdot \frac{\delta\phi^{}_{-k}}{\delta\phi^{}_x} \cdot \frac{F[\phi^{}_x]}{\Lambda^2}\right) \nonumber \\
&=& \Gamma^{}_{\rm L,UV}\left[\phi\right]+\int{ \frac{{\rm d}^4 k}{(2\pi)^4}\left(\frac{\delta\Gamma^{}_{\rm L,UV}\left[\phi\right]}{\delta\phi^{}_{-k}} \cdot \frac{F[\phi^{}_k]}{\Lambda^2}\right)}  \nonumber\\
&\equiv&\Gamma^{}_{\rm L,UV}\left[\phi\right]+\Gamma_{\rm L,UV}^\prime\left[\phi\right] \;,
\label{eq:redefiL}
\eea
where in the third line $F[\phi^{}_k] \equiv \int{{\rm d}^4 x} e^{-{\rm i}kx}F[\phi^{}_x]$ has been defined and the relation $\delta \phi^{}_{-k}/\delta \phi^{}_x = e^{-{\rm i}kx}$ has been used. In the subsequent discussions, we mainly work in the momentum space. For the moment, the on-shell condition is not implemented, so the term $\delta\Gamma^{}_{\rm L,UV}\left[\phi\right]/\delta\phi^{}_{-k}$ doesn't vanish.

After the field redefinition of $\phi$, we then focus on the $n$-point physical process with the on-shell constraints, namely,
\bea
\frac{{\rm i}\delta^{n} \Gamma^{}_{\rm L,UV}\left[\phi + F[\phi]/\Lambda^2\right]}{\delta\phi^{}_{p^{}_1} \ldots \delta\phi^{}_{p^{}_n}}\bigg|_\text{hard, on-shell} = \frac{{\rm  i}\delta^{n} }{\delta\phi^{}_{p^{}_1} \ldots \delta\phi^{}_{p^{}_n}} \int{{\rm d}^4 x} \left(\mc{L}_{\rm EFT}^{(0)}\left[\phi\right] + \mc{L}_{\rm EFT}^{(1)}\left[\phi\right]\right) \bigg|_{\rm physical} \;,
\label{eq:nleg2}
\eea
where the operators in the EFT Lagrangian on the r.h.s. belong to the physical basis instead of the Green's basis. Inserting Eq.~\eqref{eq:redefiL} into the l.h.s. of Eq.~\eqref{eq:nleg2}, the latter is divided into two parts. The first part is the derivative on $\Gamma^{}_{\rm L,UV}\left[\phi\right]$ and the second one is on $\Gamma_{\rm L,UV}^\prime\left[\phi\right]$. We shall simply denote them as $\delta^{n}\Gamma^{}_{\rm L,UV}$ and $\delta^{n}\Gamma_{\rm L,UV}^\prime$, respectively, and examine them separately.
\begin{itemize}
	\item Without imposing the on-shell conditions, we can see that $\delta^{n}\Gamma_{\rm L,UV}$ just reproduces Eq.~\eqref{eq:nleg},
	which matches with the operators in the Green's basis and thus is redundant. However, after setting the on-shell constraints, the derivative on the redundant operator will give rise to $\delta^{n} (F[\phi] \partial^2\phi)\sim p_i^2 {\cal M}(p^{}_i\cdot p^{}_j)\to -\hat{\Sigma}(\hat{m}^2) {\cal M}(p^{}_i\cdot p^{}_j)$, where $\hat{m}$ is the pole mass and $m^{}_0 = 0$ for the massless particle is assumed, and ${\cal M}(p^{}_i \cdot p^{}_j)$ represents the remaining part of the amplitude as the function of the momentum invariants $p^{}_i\cdot p^{}_j$. Such a contribution can obviously be generated by the operator $F[\phi]\phi$ in the physical basis. Under this circumstance, the on-shell conditions ensure that the $n$-point amplitude $\delta^{n}\Gamma^{}_{\rm L,UV}$ can be matched with the amplitude from non-redundant operators, and thus contribute to the coefficients of the latter.
	
	\item As for the second part $\delta^{n}\Gamma_{\rm L,UV}^\prime\left[\phi\right]$, we can explicitly figure out the variational derivative of the second term in the third line of Eq.~(\ref{eq:redefiL}), i.e.,
	\bea
	\frac{{\rm i}\delta^{n}\Gamma_{\rm L,UV}^\prime\left[\phi\right]}{\delta\phi^{}_{p^{}_1} \ldots \delta\phi^{}_{p^{}_n}}\bigg|_\text{hard, on-shell} &=& \int\frac{{\rm d}^4 k}{\left(2\pi\right)^4} \left\{\frac{{\rm i}\delta\Gamma^{}_{\rm L,UV}\left[\phi\right]}{\delta\phi^{}_{-k}} \cdot \frac{1}{\Lambda^2} \frac{\delta^{n}F\left[\phi^{}_k\right]}{\delta\phi^{}_{p^{}_1} \ldots \delta\phi^{}_{p^{}_n}} \right. \nonumber \\
	&& +\sum_{i=1}^{n} \frac{{\rm i}\delta^{2}\Gamma^{}_{\rm L,UV}\left[\phi\right]}{\delta\phi^{}_{p^{}_i}\delta\phi^{}_{-k}} \cdot\frac{1}{\Lambda^2} \frac{\delta^{n-1}F\left[\phi^{}_k\right]}{\delta\phi^{}_{p^{}_1} \ldots \delta \xcancel{\phi^{}_{p^{}_i}} \ldots \delta\phi^{}_{p^{}_n}} \nonumber \\
	&& \left. +\sum_{I,J>1}^{I+J=n} { \frac{{\rm i} \delta^{I+1}\Gamma^{}_{\rm L,UV}\left[\phi\right]}{\delta\phi^{}_{p^{}_1} \ldots \delta\phi^{}_{p^{}_I} \delta\phi^{}_{-k}} \cdot\frac{1}{\Lambda^2}\frac{\delta^{J} F\left[\phi^{}_k \right]}{\delta\phi^{}_{p^{}_{I+1}} \ldots \delta\phi^{}_{p^{}_{n}}}  }\right\} \bigg|_\text{hard, on-shell}\;, \qquad
	\label{eq:secondterm}
	\eea
	where on the r.h.s. we have exhausted all possibilities to take the $n$-th-order derivative of $F[\phi]$ and $\Gamma^{}_{\rm L, UV}[\phi]$. In the first line, all derivatives act on $F[\phi]$, while in the second line all derivatives act on $F[\phi]$ except for one acting on $\Gamma^{}_{\rm L, UV}[\phi]$, and so on. After imposing the on-shell conditions in Eqs.~\eqref{eq:eom} and \eqref{eq:onshell}, one can easily see that the terms in the first and second lines vanish, but those in the third line survive.
	
	To deal with the terms in the third line of Eq.~(\ref{eq:secondterm}), we can make use of the following identity
	\begin{eqnarray}
		{\bf 1} = -\frac{1}{k^2} \frac{\delta (\partial^2_x \phi^{}_x)}{\delta \phi^{}_k} e^{{\rm i}kx} \; , \label{eq:identity}
	\end{eqnarray}
	 which can be immediately verified by using the Fourier transformation of the field $\phi^{}_x$ in the coordinate space to $\phi^{}_k$ in the momentum space or equivalently $\delta\phi^{}_x/\delta\phi^{}_k = e^{-{\rm i}kx}$. After inserting the identity in Eq.~(\ref{eq:identity}) into the third line of Eq.~\eqref{eq:secondterm}, one gets
	\bea
	&&\frac{{\rm i}\delta^{I+1}\Gamma^{}_{\rm L,UV}\left[\phi\right]}{\delta\phi{}_{p^{}_1} \ldots \delta\phi^{}_{p^{}_I} \delta\phi^{}_{-k}} \cdot \frac{\delta^{J}}{\delta\phi^{}_{p^{}_{I+1}}\ldots\delta\phi^{}_{p^{}_n}} \left( \int{{\rm d}^4x} e^{-{\rm i}kx}   \frac{F\left[\phi^{}_x\right]}{\Lambda^2}\right)\bigg|_\text{hard, on-shell}\nonumber \\
	&=&\frac{{\rm i}\delta^{I+1}\Gamma^{}_{\rm L,UV}\left[\phi\right]}{\delta\phi^{}_{p^{}_1} \ldots \delta\phi^{}_{p^{}_i} \delta\phi^{}_{-k}} \cdot \frac{\delta^{J}}{\delta\phi^{}_{p^{}_{I+1}}\ldots\delta\phi^{}_{p^{}_n}} \left(\int{{\rm d}^4x}\frac{-1}{k^2} \frac{\delta(\partial^2_x\phi^{}_x)}{\delta\phi^{}_k} \frac{F\left[\phi^{}_x\right]}{\Lambda^2} \right) \bigg|_\text{hard, on-shell} \nonumber \\
	&=& \frac{{\rm i}\delta^{I+1}\Gamma^{}_{\rm L,UV}\left[\phi\right]}{\delta\phi^{}_{p^{}_1} \ldots \delta\phi^{}_{p^{}_I}\delta\phi^{}_{-k}} \cdot \frac{\rm i}{k^2} \cdot \frac{{\rm i}\delta^{J+1}}{\delta\phi^{}_{p^{}_{I+1}} \ldots \delta\phi^{}_{p^{}_n} \delta\phi^{}_k}\left(\int{{\rm d}^4x}\frac{ F\left[\phi^{}_x\right] \partial^2_x\phi^{}_x }{\Lambda^2} \right) \bigg|^{\text{take}\ (k^2)^0 }_\text{hard, on-shell}  \nonumber \\
	&=& \frac{{\rm i}\delta^{I+1}\Gamma^{}_{\rm L,UV}\left[\phi\right]}{\delta\phi^{}_{p^{}_1} \ldots \delta\phi^{}_{p^{}_I}\delta\phi^{}_{-k}} \cdot \frac{\rm i}{k^2} \cdot \frac{{\rm i}\delta^{J+1} \Gamma^{}_{\rm L,UV} \left[\phi\right] }{\delta\phi^{}_{p^{}_{I+1}} \ldots \delta\phi^{}_{p^{}_n} \delta\phi^{}_k} \bigg|^{\text{take}\ (k^2)^0 }_{\text{hard, on-shell}}\;,
	\label{eq:trick}
	\eea
where the identity has been utilized in the second line. Some explanations for the derivation and notations in Eq.~(\ref{eq:trick}) are necessary.
	\begin{enumerate}
		\item From the second to third line, the variation $\delta/\delta\phi^{}_k$ has been extended from $\partial^2 \phi$ to the whole operator $F[\phi]\partial^2\phi$. Such an extension is justified for two reasons. First, $\delta F[\phi]/\delta\phi^{}_k$ does not provide any terms proportional to the momentum $k^2$ or its higher power. This is guaranteed by the requirement that $F[\phi]$ itself cannot be redundant as previously stated, i.e., it does not consist of the term $\partial^2\phi$. Second, the $(I+1)$-point vertex $\delta^{I+1}\Gamma^{}_{\rm L,UV}$ is only a function of the scalar products ${p^{}_i\cdot p^{}_j}$ or ${p^{}_i\cdot k}$, so it cannot give a factor of $k^2$ as well.\footnote{In the effective action $\Gamma^{}_{\rm L,UV}$, only the kinetic term $-\phi\partial^2\phi$ and $F\partial^2\phi/\Lambda^2$ can provide a factor of $k^2$. However, the kinetic term only contributes to the 2-point amplitude, which has been forbidden by the constraint $I > 1$. As for the $F\partial^2\phi/\Lambda^2$ term, when combined with another term of $F\partial^2\phi/\Lambda^2$, it is of higher order of $1/\Lambda^2$ and thus can be ignored.} We are concerned about the $k^2$ factor, which will be able to cancel the propagator $1/k^2$ out, because any amplitudes with the momentum $k$ (and thus the external momenta) in the denominator cannot be produced by local operators. In the end, it has to be the $\partial^2\phi$ that cancels out the $1/k^2$ factor, leading to nontrivial contributions to the amplitudes that reproduced by the operators in the physical basis. All these arguments explain the meaning of the superscript ``take $(k^2)^0$'' in the third line of Eq.~\eqref{eq:trick}. That is to set the constraint that only the amplitudes of order $(k^2)^0$ are retained.
		
		\item For the third to fourth line of Eq.~\eqref{eq:trick}, we reexpress the operator $F[\phi]\partial^2\phi$ in terms of the effective action $\Gamma^{}_\text{L,UV}[\phi]$ according to Eq.~\eqref{eq:nleg}. Since the hard-region and the on-shell conditions have been required all along, the locality of $\Gamma^{}_{\rm L,UV}[\phi]$ is ensured. In this way, the formula in the fourth line of Eq.~\eqref{eq:trick} becomes formally more symmetric and physically more transparent, as we shall see below.
	\end{enumerate}
\end{itemize}

To sum up, starting with the UV full theory and its 1LPI effective action $\Gamma^{}_{\rm L, UV}$, we adopt following master equation to match the UV theory directly to the EFT in the physical basis without the redundant $F[\phi]\partial^2\phi$ operator
\bea
\left[\frac{{\rm i}\delta^{n} \Gamma^{}_{\rm L,UV}\left[\phi\right]}{\delta\phi^{}_{p^{}_1} \cdots \delta\phi^{}_{p^{}_n}} \right. + \left. \int\frac{{\rm d}^4k}{\left(2\pi\right)^4} \sum_{I,J>1}^{I+J=n} \left.\frac{{\rm i}\delta^{I+1}\Gamma_{\rm L,UV}\left[\phi\right]}{\delta\phi^{}_{p^{}_1}\cdots\delta\phi^{}_{p^{}_I}\delta\phi^{}_{-k}} \cdot \frac{\rm i}{k^2}\cdot\frac{{\rm i}\delta^{J+1} \Gamma^{}_{\rm L,UV} \left[\phi\right] }{\delta\phi^{}_{p^{}_{I+1}} \cdots\delta\phi^{}_{p^{}_n} \delta\phi^{}_k} \right] \right|^{\text{take}\ (k^2)^0 }_\text{\rm hard, on-shell}\;.
\label{eq:maineq}
\eea
It is worth commenting on the momentum conservation in Eq.~\eqref{eq:maineq}. Due to Eq.~\eqref{eq:Gamma}, the effective action  in the first term of Eq.~\eqref{eq:maineq} actually carries a global factor of momentum conservation $(2\pi)^4\delta^{4}(p^{}_1+\cdots+p^{}_n)$. For the second term, one can verify that it possesses the same factor, i.e.,
\bea
\int{\frac{{\rm d}^4 k}{(2\pi)^4}}(2\pi)^4\delta^{4}(p^{}_1+\cdots+p^{}_I-k)\cdot (2\pi)^4\delta^{4}(p^{}_{I+1}+\cdots+p^{}_n + k)=(2\pi)^4\delta^{4}(p^{}_1+\cdots+p^{}_n) \;, \nonumber
\eea
as it should. Notice that the first term in Eq.~\eqref{eq:maineq} is the traditional hard-region contribution from the $n$-point 1LPI diagrams but now constrained by the on-shell conditions, whereas the second term represents new diagrams that need to be taken into account. This corresponds to reducible diagrams that contain a tree-level light propagator ${\rm i}/k^2$ linking two 1LPI diagrams. It is reasonable to call them the two-light-particle-irreducible (2LPI) diagrams. Those two terms in Eq.~\eqref{eq:maineq} have been shown as the first and second diagrams in Fig.~\ref{maindiag}.

\subsection{Generalization}
In the previous proof, we have concentrated only on the scalar field, but it can be generalized in a straightforward way to the theory that contains fermions and gauge bosons. Several aspects of the generalization will be investigated in this subsection.
\begin{itemize}
\item {\bf Covariant derivative}. In the presence of gauge fields, to maintain the gauge invariance, we must replace the ordinary derivative with the covariant derivative, i.e., $\partial^{}_\mu \to D^{}_\mu$. Therefore, in the EFT of a scalar field $\phi$ in the nontrivial representation of the gauge symmetry group, the redundant operator becomes $F[\phi]D^2\phi$. In this case, we still insert the identity in Eq.~(\ref{eq:identity}) into Eq.~\eqref{eq:trick}. Although it is not the covariant derivative, it does not affect our conclusion because of the following relation
\bea
\int{{\rm d}^4x} \frac{-1}{k^2} \frac{\delta(\partial^2_x \phi^{}_x)}{\delta\phi^{}_k} \frac{F\left[\phi^{}_x\right]}{\Lambda^2} = \frac{-1}{k^2} \frac{\delta}{\delta\phi^{}_k} \left( \int{{\rm d}^4x} \frac{F\left[\phi^{}_x\right]D^2_x \phi^{}_x}{\Lambda^2} \right)\bigg|^{\text{take }(k^2)^0}\;,
\label{eq:covar}
\eea
where $\partial^2$ has been safely replaced by $D^2$. The validity of Eq.~\eqref{eq:covar} is evident by noticing that the difference between $\partial^2\phi$ and $D^2\phi$ doesn't bring in any $k^2$ factor, and they are actually equivalent when the constraint of ``take $(k^2)^0$'' is set.

There might be a potential problem. The first term of Eq.~\eqref{eq:maineq}, i.e., $\delta^{n}\Gamma^{}_{\rm L,UV}$, leads to the redundant operator $F[\phi]D^2\phi$, which turns out to be non-redundant because of $p_i^2\to -\hat{\Sigma}(\hat{m}^2)$ for the on-shell 1LPI $n$-$\phi$ vertex. However, when we are dealing with the $n$-point vertex with one leg being a gauge boson, such a redundant operator seems impossible to be converted into that in the physical basis even with the on-shell conditions imposed. More explicitly,
recalling the definition $D^2\phi=\left(\partial_\mu-igA_\mu\right)^2\phi$ and the on-shell conditions, we have
\bea
\int{{\rm d}^4 x}\frac{{\rm i}\delta^n \left( F\left[\phi^{}_x\right] D^2\phi^{}_x \right)}{\delta\phi^{}_{p^{}_1} \cdots \delta\phi^{}_{p^{}_{n-1}} \delta A_{p^{}_n}^\mu}  \supset \int{{\rm d}^4 x} \frac{{\rm i}\delta^{n-2} F\left[\phi^{}_x\right]}{\delta\phi^{}_{p^{}_1} \cdots \delta \xcancel{ \phi^{}_{p^{}_i}} \cdots \delta\phi^{}_{p^{}_{n-1}}}  \frac{\delta^2(D^2\phi^{}_x)}{\delta\phi_{p_i}\delta A_{p_n}^\mu}
\sim   p_i\cdot\epsilon\left(p_n\right)\neq0 \;, \quad
\label{eq:vectorleg}
\eea
where $\epsilon^\mu(p^{}_n)$ denotes the polarization vector of the gauge boson. As the term in Eq.~\eqref{eq:vectorleg} is non-vanishing and seemly needs to be matched only by the operator the Green's basis. Fortunately, we can prove that this term can be offset by a contribution from the second term of Eq.~\eqref{eq:maineq}, which consists of
\bea
\int{\frac{{\rm d}^4k}{(2\pi)^4}}\frac{{\rm i}\delta^{3}\Gamma^{}_{\rm L,UV}\left[\phi\right]}{\delta\phi^{}_{p^{}_i}\delta A_{p^{}_n}^\mu \delta\phi^{}_{-k}} \frac{\delta^{n-2} F[\phi^{}_k]}{\delta\phi^{}_{p^{}_1} \cdots \delta \xcancel{\phi^{}_{p^{}_i}} \cdots \delta\phi^{}_{p^{}_{n-1}}}  \supset  - \int{{\rm d}^4x} \frac{{\rm i}\delta^2 (D^2\phi^{}_x)}{\delta\phi^{}_{p^{}_i}\delta A_{p^{}_n}^\mu} \frac{\delta^{n-2} F\left[\phi^{}_x\right]} {\delta\phi^{}_{p^{}_1}\cdots\delta\xcancel{\phi^{}_{p^{}_i}} \cdots \delta\phi^{}_{p^{}_{n-1}}}\;, \nonumber \\
\label{eq:vectorleg2}
\eea
where we single out the kinetic term $(-1/2) \int{{\rm d}^4x}\; \phi^{}_x D^2\phi^{}_x$ from $\Gamma^{}_{\rm L,UV}$. It is interesting to observe that the result in Eq.~\eqref{eq:vectorleg} just cancels out that in Eq.~\eqref{eq:vectorleg2}. That is, the second term in Eq.~\eqref{eq:maineq} can offset the redundant-operator contributions of the first term. Therefore, as long as we calculate all possible diagrams in Fig.~\ref{maindiag}, it is sufficient to absorb all on-shell amplitudes into the physical basis without introducing the Green's basis.

\item {\bf Field redefinition}. The field redefinitions of fermions and gauge bosons are similar to that of the scalar. For example, the redundant operator for fermions takes the form of $F\slashed{D} \psi/\Lambda^2$. In this case, the field redefinition should be $\overline{\psi} \rightarrow \overline{\psi} + F/\Lambda^2$~\cite{Arzt:1993gz}. Then, one can follow the same arguments leading to Eq.~\eqref{eq:secondterm} as in the scalar case, then we should insert another new but similar identity
    \begin{eqnarray}
    {\bf 1} = -\frac{1}{\slashed{k}} \frac{\delta (\slashed{\partial}^{}_x\psi^{}_x)}{\delta\psi^{}_k }e^{{\rm i}kx} \; ,
    \label{eq:newidentity}
    \end{eqnarray}
     into Eq.~(\ref{eq:trick}). The rest way to the final conclusion is almost the same as before. In the case of gauge bosons, the form of the redundant operator will be $F_a^\nu D^\mu A_{\mu\nu}^a/\Lambda^2$, where $A^a_{\mu\nu}$ stands for the field-strength tensor with $a$ being the group-generator index. It is easy to find the field redefinition $A_a^\nu \rightarrow A_a^\nu - F_a^\nu/\Lambda^2$~\cite{Arzt:1993gz}, and to reach the conclusion in Eq.~\eqref{eq:maineq}.

     \item {\bf Higher-order derivatives}. Thus far we have investigated the case where the effective operator can be reduced by applying the EOM once, such as $FD^2\phi$. Nevertheless, there is another type of effective operator that needs to be removed by applying the EOM twice, e.g., $(D^{}_\mu A^{\mu\rho})(D^\nu A^{}_{\nu\rho})$ at dim-6. We have to calculate the 3LPI diagrams to offset such a redundant operator, and the Feynman diagrams look like
\bea
\begin{gathered}
        \includegraphics[width=.35\linewidth]{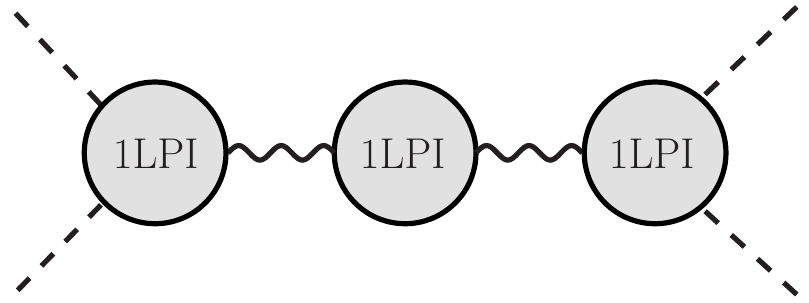}
\end{gathered}\nonumber
\eea
This can be understood as follows. The operator $(D^{}_\mu A^{\mu\rho})(D^\nu A^{}_{\nu\rho})$ produces two $k^2$ factors, which need to be canceled out by two light $A^a_\mu$ propagators.

When we turn to higher-dimensional operators in the EFT, there can be an operator that contains $m$ copies of $D^2\phi$ to be reduced by the EOM $m$ times, like $F(D^2\phi)^m$. Therefore, we must calculate the $m$-LPI diagrams, in which $m$ light propagators that need to be cut to make it disconnected, c.f. the last diagram in Fig.~\ref{maindiag}. In practice, we just take account of all possible diagrams that contribute to this process. If the number of light propagators of such kind in a diagram is greater than $m$, it will be irrelevant. This is because the power of momentum in the denominator of this diagram is higher than that in $F(D^2\phi)^m$, and should be ignored according to the rule ``retaining only those terms whose denominators do not contain external momenta.''

\item {\bf Particle masses}. When the light particle $\phi$ is massive, the proof can be slightly modified accordingly. First, the first term in Eq.~\eqref{eq:maineq} implies the derivative of a redundant operator $\delta^n\left(FD^2\phi\right) \sim p_i^2\rightarrow m^2_0 -\hat{\Sigma}(\hat{m}^2)$ under the one-shell conditions. This does not change the conclusion that its contribution can be compensated by the operators in the EFT in the physical basis. Then, the identity inserted into Eq.~\eqref{eq:trick} changes to
    \begin{eqnarray}
    {\bf 1} = \frac{-1}{k^2-\hat{m}^2} \frac{\delta \left[\left(\partial^2_x + \hat{m}^2\right)\phi^{}_x\right]}{\delta\phi^{}_k} e^{{\rm i}k x} \; .
    \end{eqnarray}
    As a result, Eq.~\eqref{eq:trick} will be modified by adding an extra term
\bea
\frac{{\rm i}\delta^{I+1}\Gamma^{}_{\rm L,UV}\left[\phi\right]}{\delta\phi^{}_{p^{}_1} \ldots \delta\phi^{}_{p^{}_I}\delta\phi^{}_{-k}} \cdot \frac{\rm i}{k^2 - \hat{m}^2} \cdot \frac{\rm i}{\Lambda^2} \frac{\hat{m}^2 \delta^{J+1}}{\delta\phi^{}_{p^{}_{I+1}} \ldots \delta\phi^{}_{p^{}_n}\delta\phi^{}_k} \int{{\rm d}^4x \left(F[\phi^{}_x]\phi^{}_x\right)}\;, \nonumber
\eea
which disappears after imposing the condition ``take $(k^2-\hat{m}^2)^0$.'' Hence the conclusion in Eq.~\eqref{eq:maineq} remains valid.

\end{itemize}

In the above cases, we don't see any difficulty in generalizing our proof to the theories with scalar bosons, fermions and gauge bosons. Hence the on-shell-amplitude approach presented in Sec.~\ref{sec:convention} can be applied to any general field theories.

\section{Applications}
\label{sec:application}
\subsection{One-loop matching of a toy model}
As a specific example for one-loop matching, we will once again consider the simple UV model that includes both a light scalar $\phi$ and a heavy scalar $\Phi$. This model has previously been studied in Refs.~\cite{Henning:2016lyp, Fuentes-Martin:2016uol} and more recently in Ref.~\cite{DeAngelis:2023bmd}. However, as we have already pointed out, Refs.~\cite{Henning:2016lyp,Fuentes-Martin:2016uol} calculate the 1LPI off-shell amplitudes, which are matched to the operators in the Green's basis at the one-loop level. In this subsection, we compute the on-shell amplitudes and directly match them to the operators in the physical basis. As a cross-check, we compare our results with those in Ref.~\cite{Fuentes-Martin:2016uol}, where the redundant operators are removed by using the EOM, and find an excellent agreement with each other.

The UV model is described by the following Lagrangian
\bea
\mc{L}\left[\Phi,\phi\right] = \frac{1}{2}\left(\partial^{}_\mu\phi \partial^\mu\phi - m^2\phi^2\right) + \frac{1}{2}\left(\partial^{}_\mu\Phi \partial^\mu\Phi - M^2\Phi^2\right) - \frac{\kappa}{4!}\phi^4 - \frac{\lambda}{3!}\phi^3\Phi \;,
\label{eq:UVtoy}
\eea
where the light scalar mass $m$ and the heavy scalar mass $M$ satisfy the condition $m \ll M$, and $\{\kappa, \lambda\}$ are dimensionless coupling constants. As shown in Ref.~\cite{Fuentes-Martin:2016uol}, using the conventional method of one-loop matching, the redundant operators appear in the EFT Lagrangian, namely,
\bea
\mc{L}^{}_{\rm EFT} = \frac{1}{2}\left(\partial^{}_\mu\phi \partial^\mu\phi - m^2\phi^2\right) - \frac{\kappa}{4!} \phi^4 + \frac{\lambda^2}{72M^2}\phi^6 + \frac{\alpha}{4!}\phi^4 + \frac{\beta}{4!M^2}\phi^2 \partial^2\phi^2 + \frac{\gamma}{6!M^2}\phi^6 \;,
\label{eq:redun-scalar}
\eea
where the operators associated with the coefficients $\{\alpha, \beta, \gamma\}$ arise from the one-loop matching and their matching conditions are
\bea
\alpha=\frac{3\lambda^2}{16\pi^2}\left(1+\frac{m^2}{M^2}\right),\quad \beta=-\frac{3\lambda^2}{16\pi^2}\frac{1}{2},\quad \gamma=\frac{45}{16\pi^2}\kappa\lambda^2 \;.
\eea
The second operator $\phi^2\partial^2\phi^2$ from the one-loop matching is redundant and can be reexpressed as $\phi^2\partial^2\phi^2 = (4/3) \phi^3\partial^2\phi$ by using the relation from integration by parts.
For this toy model, we regard the basis with redundant operators in Eq.~(\ref{eq:redun-scalar}) as the Green's basis, and that with only two independent operators $\phi^4$ and $\phi^6$ at dim-4 and dim-6, respectively, as the physical basis.

Then we make use of the on-shell-amplitude approach to directly derive the coefficients of the physical operators $\phi^4$ and $\phi^6$. Starting with the Lagrangian of the UV model in Eq.~(\ref{eq:UVtoy}), one should calculate on-shell amplitudes of all possible diagrams, and show that they can be absorbed into the coefficients of $\phi^4$ and $\phi^6$. At this point, we stress that the physical basis of independent and complete operators up to a given mass dimension in the EFT should be known in the first place. This is usually the case when matching the UV theory onto the SMEFT. In the toy model under consideration, two physical operators produce two processes $\phi+\phi\to \phi+\phi $ and $\phi+\phi+\phi\to \phi+\phi+\phi$ at the tree level. Therefore, we calculate the Feynman diagrams for these two processes in the UV theory at the one-loop level, and extract only the hard-region contributions and on-shell amplitudes. The relevant diagrams and results can be found below.
\bea
\begin{gathered}
	\includegraphics[width=.23\linewidth]{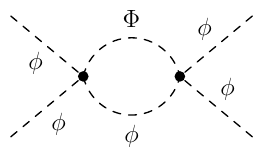}
\end{gathered}
&=&\frac{3\lambda^2}{16\pi^2}+\frac{5m^2}{M^2}\frac{\lambda^2}{16\pi^2}\;, \nonumber \\
\begin{gathered}
	\includegraphics[width=.23\linewidth]{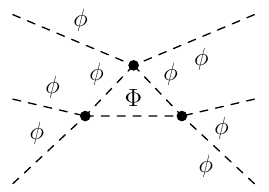}
\end{gathered}
&=& \frac{45}{16\pi^2M^2}\kappa\lambda^2 \;, \nonumber \\
\begin{gathered}
	\includegraphics[width=.23\linewidth]{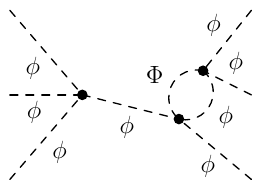}
\end{gathered}
&=& \frac{5\kappa\lambda^2}{8\pi^2M^2} + \left(\begin{matrix} \text{terms with momentum}\\\text{in the denominators}\\\end{matrix}\right) \;.
\eea
The first diagram is irreducible and contributes to the coefficient of $\phi^4$. The second and third diagrams contribute to that of $\phi^6$, where the third one is reducible and therefore contains some terms having the momentum in the denominators. As we have stated in Sec.~\ref{sec:proof}, such terms have to be dropped in our matching procedure. Finally, we obtain the physical EFT Lagrangian
\bea
\mc{L}_{\rm EFT}
&=&\frac{1}{2}\left(\partial_\mu\phi\partial^\mu\phi-m^2\phi^2\right)-\frac{1}{4!}\left(\kappa-\frac{3\lambda^2}{16\pi^2}-\frac{5m^2}{M^2}\frac{\lambda^2}{16\pi^2}\right)\phi^4+\frac{1}{6!M^2}\left(10\lambda^2+ \frac{55\kappa\lambda^2}{16\pi^2} \right)\phi^6 \;. \qquad
\label{eq:scalarmatching}
\eea

By manually applying the EOM, i.e, $\partial^2\phi = -m^2 \phi - \left(\kappa + \alpha\right)\phi^3/3!$, to Eq.~\eqref{eq:redun-scalar}, one can verify that the results in Eq.~\eqref{eq:scalarmatching} can also be derived. However, the on-shell-amplitude approach is more direct and efficient than the conventional one used in the previous literature~\cite{Henning:2016lyp,Fuentes-Martin:2016uol}. Apart from this simple UV model, a similar strategy has been adopted in Ref.~\cite{Li:2022aby} to accomplish the one-loop matching of more complicated UV models onto the dimension-eight physical basis. These examples include the scalar extensions of the SM by a heavy doublet with $Y = 1/2$, or by heavy quadruplets with $Y = 1/2$ and $3/2$. The results are found to be in perfect agreement with the ones based on the off-shell matching method~\cite{Chala:2021wpj}.

\subsection{Beta functions in the SEFT-I}
Although the approach introduced in Sec.~\ref{sec:proof} is intended for the one-loop matching of the UV full theory to the EFT, it is not difficult to find that the proof is not limited to whether the loop diagram is calculated in the UV theory or the EFT. In other words, we can use the generating functional $\Gamma^{}_{\rm EFT}[\phi]$, instead of $\Gamma^{}_{\rm L,UV}[\phi]$, to calculate the loop diagram with one insertion of the operator in the EFT. The divergent part of the loop diagram will be absorbed into the counterterms of EFT operators. However, the traditional method uses the operators in the Green's basis to do so, and then converts these counterterms to those in the physical basis by the EOM. Interestingly, based on the method given in Sec.~\ref{sec:proof}, we are able to directly obtain the counterterms in the physical basis by implementing the field redefinition in $\Gamma^{}_{\rm EFT}[\phi]$. The proof will be similar to that given in Sec.~\ref{sec:proof}, except that we replace $\Gamma^{}_{\rm L,UV}[\phi]$ by $\Gamma^{}_{\rm EFT}[\phi]$ and drop the ``$|_{\rm hard}$ '' constraint consistently. This is safe because the hard-region requirement just ensures the locality of $\Gamma^{}_{\rm L,UV}[\phi]$. Under the present circumstance, the locality of $\Gamma^{}_{\rm EFT}[\phi]$ is guaranteed since only the UV divergence of the loop diagrams are of our concern. Hence, we will not elaborate on the proof further.
\begin{figure}[t]
	\centering
	\includegraphics[width=1.\linewidth]{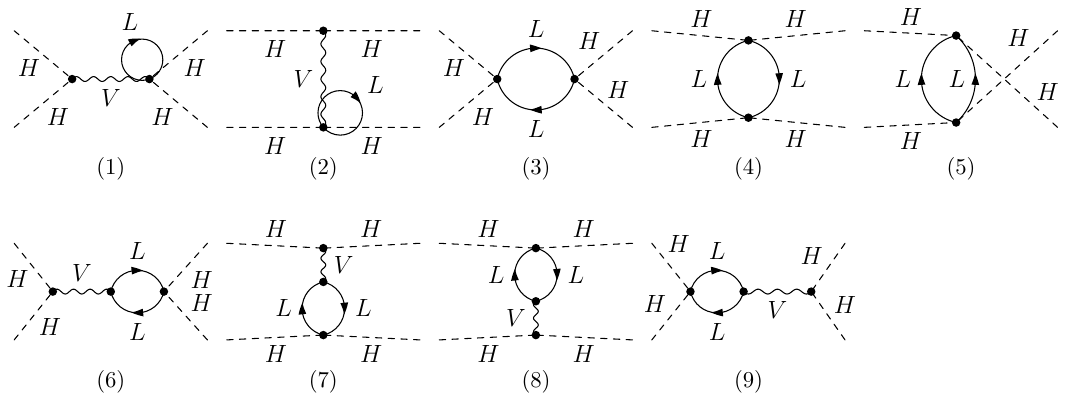}
	\caption{All possible one-loop diagrams in the SEFT-I that contribute to $\phi^{}_1 \overline{\phi}^{}_1 \to \phi^{}_1 \overline{\phi}^{}_1$, where $V \in \{B,W_I\}$ denote the ${\rm U}(1)^{}_{\rm Y}$ and ${\rm SU}(2)^{}_{\rm L}$ gauge fields and $L$ is the lepton doublet.}
	\label{beta-diag}
\end{figure}

In this subsection, by a concrete example, we explain how to directly derive the beta functions for the Wilson coefficients of the operators in the physical basis. The definition of the beta function for the Wilson coefficient $c^i_6$ associated with the dim-6 operator $O_6^i$ is given by
\bea
\beta\left(c_6^i\right) \equiv 16\pi^2 \mu\frac{{\rm d} c_6^i}{{\rm d}\mu} = \gamma^{}_{ij} c_6^j + \gamma^\prime (c^{}_5)^2 \;,
\label{eq:definitionbeta}
\eea
where $\gamma^{}_{ij}$ represents the anomalous dimension of the single insertion of a dim-6 operator $O_6^j$, while $\gamma^\prime$ denotes that of the double insertions of the unique dim-5 operator $O^{}_5$. Unlike the one-loop matching, we need to compute the EFT one-loop amplitudes for the r.h.s. of Eq.~\eqref{eq:definitionbeta} and associate them with the EFT tree-level amplitudes on the l.h.s. We emphasize that the operators in question all belong to the physical basis.

To be specific, we work in the type-I seesaw effective field theory
(SEFT-I). The SEFT-I extends the SM by three operators $O^{}_5$, $O_{Hl}^{\left(1\right)}$, and $O_{Hl}^{\left(3\right)}$. They are obtained by integrating out heavy right-handed neutrinos at the tree level. In addition, for simplicity, we focus on 4-$H$ processes, and calculate the beta functions for relevant dim-6 operators caused by the operators in the SEFT-I. Here the Higgs doublet is $H=\left(\phi^{}_1, \phi^{}_2\right)^{\rm T}$ and its Hermitian conjugate is $H^\dag=\left(\phi_1^\ast,\phi_2^\ast\right)$.

On the l.h.s. of Eq.~\eqref{eq:definitionbeta}, there will be three operators involved in the 4-Higgs processes at the tree level, namely,
$\lambda\left(H^\dag H\right)^2$, $C^{}_{HD}O^{}_{HD}$ and $C^{}_{H\square } O^{}_{H\square}$. We choose three unrelated processes $\phi^{}_1 \overline{\phi}^{}_1 \rightarrow \phi^{}_1 \overline{\phi}^{}_1$, $\phi^{}_1\phi^{}_2 \rightarrow \phi^{}_1\phi^{}_2$ ($t$-channel) and $\phi^{}_1\phi^{}_2\rightarrow\phi^{}_1\phi^{}_2$ ($u$-channel), whose tree-level amplitudes in terms of $\lambda$, $C^{}_{HD}$ and $C_{H\square }$ are respectively given by
\bea
{\cal M}_{1}^{(0)}&=&\frac{ C^{}_{HD}\left(4m_h^2-t-s\right)}{\mathrm{\Lambda}^2}-\frac{2C^{}_{H\square}(s+t)}{\mathrm{\Lambda}^2}-4\lambda \;, \label{eq:betac1}\\
{\cal M}_{2}^{(0)}&=&\frac{C^{}_{HD}\left(s+t-2m_h^2\right) }{\mathrm{\Lambda}^2}-\frac{2C^{}_{H\square}t}{\mathrm{\Lambda}^2}-2\lambda \;, \\
{\cal M}_{3}^{(0)} &=&\frac{C^{}_{HD}(2m_h^2-t)}{\mathrm{\Lambda}^2}+\frac{2C^{}_{H\square}(s+t-4m_h^2)}{\mathrm{\Lambda}^2}-2\lambda \;.
\label{eq:betac3}
\eea
Note that the mass $m^{}_h$ of the external Higgs boson is kept in the above equations, and $\{s, t\}$ denote the Mandelstam variables.

On the r.h.s of Eq.~\eqref{eq:definitionbeta}, we have to calculate all one-loop diagrams in the SEFT-I, which are shown in Fig.~\ref{beta-diag}. After taking on-shell conditions and dropping  terms containing momenta in the denominator, we obtain the divergent parts of the amplitudes for three processes
\bea
{\cal M}^{}_1|^{}_{\rm div}&=& \frac{ (4m_h^2-s-t)}{8\pi^2} \tr\left[C_5^{} C_5^\dag\right]-\frac{e^2\left(8m_h^2-3\left(s+t\right)\right)}{24\pi^2c_{\rm W}^2  s_{\rm W}^2} (\tr\left[C_{Hl}^{\left(3\right)}\right] c_{\rm W}^2 - \tr\left[C_{Hl}^{\left(1\right)}\right]s_{\rm W}^2)  \;, \label{eq:gammac1}\\
{\cal M}^{}_2|^{}_{\rm div}&=& \frac{s}{16\pi^2} \tr\left[C_5^{} C_5^\dag\right]+\frac{e^2\tr\left[C_{Hl}^{\left(1\right)}\right] s_{\rm W}^2 \left(2s+t-4m_h^2\right) +e^2 \tr\left[C_{Hl}^{\left(3\right)}\right]c_{\rm W}^2 \left(3t-4m_h^2\right)}{24\pi^2 c_{\rm W}^2 s_{\rm W}^2} \;, \\
{\cal M}^{}_3|^{}_{\rm div}&=& \frac{s}{16\pi^2}\tr\left[C^{}_5 C_5^\dag\right]+\frac{e^2 \tr\left[C_{Hl}^{\left(1\right)}\right] s_{\rm W}^2 \left(s-t\right) + e^2\tr\left[C_{Hl}^{\left(3\right)}\right] c_{\rm W}^2 \left(8m_h^2-3\left(s+t\right)\right)}{24\pi^2c_{\rm W}^2 s_{\rm W}^2} \;,
\label{eq:gammac3}
\eea
where $c^{}_{\rm W} \equiv \cos \theta^{}_{\rm W}$ and $s^{}_{\rm W} \equiv \sin \theta^{}_{\rm W}$ with $\theta^{}_{\rm W}$ being the weak mixing angle, and $e$ is the electromagnetic gauge coupling. By combining Eqs.~\eqref{eq:betac1}-\eqref{eq:betac3} with Eqs.~\eqref{eq:gammac1}-\eqref{eq:gammac3}, and recalling that these divergences are to be absorbed by the operators in the physical basis, we get the beta functions as follows
\bea
\beta\left(C^{}_{HD}\right)&=&-2\tr\left[C^{}_5 C_5^\dag\right]-\frac{8}{3}g_1^2\tr\left[C_{Hl}^{\left(1\right)}\right] \;, \nonumber \\
\beta\left(C^{}_{H\square}\right)&=&-\tr\left[C^{}_5 C_5^\dag\right]-2e^2
\left(\frac{1}{3}g_1^2\tr\left[C_{Hl}^{\left(1\right)}\right]-g_2^2\tr\left[C_{Hl}^{\left(3\right)}\right] \right)  \;,\nonumber \\
\beta\left(\lambda\right)&=&2m_h^2\left(\tr\left[C^{}_5 C_5^\dag\right]-\frac{4}{3}g_2^2\tr\left[C_{Hl}^{\left(3\right)}\right]\right) \;, \label{eq:betaSEFTI}
\eea
where $g^{}_1$ and $g^{}_2$ are the ${\rm U}(1)^{}_{\rm Y}$ and ${\rm SU}(2)^{}_{\rm L}$ gauge couplings. One can verify that Eq.~\eqref{eq:betaSEFTI} reproduces exactly the results in Refs.~\cite{Jenkins:2013zja, Broncano:2004tz, Davidson:2018zuo, Wang:2023bdw}. Note also that the anomalous dimensions are independent of the explicit expressions of $C^{}_5$, $C_{Hl}^{\left(1\right)}$ and $C_{Hl}^{\left(3\right)}$, so this method can be used in a more general EFT, apart from the specific one SEFT-I.

\subsection{Evanescent contributions in the SEFT-II}

In the EFT, the Fierz identities are frequently used to transform the operators from one type to another. Howerver, these identities are exactly true only in the four-dimensional spacetime. Hence the the Fierz identities for some operators may result in additional terms proportional to $\varepsilon = 4-d$ in the dimensional regularization with $d$ being the spacetime dimension. These operators automatically disappear in the four-dimensional spacetime, i.e., $\varepsilon \to 0$. Nevertheless, when such an operator is inserted into a loop diagram, the UV-divergence $1/\varepsilon$ in the loop calculation will cancel out the coefficient $\varepsilon$ of this operator, resulting in a finite term that does not necessarily vanish in dim-4. As these contributions are of one-loop order, they can be absorbed into the one-loop matching coefficients of the operators, because of the completeness of the physical basis.

In this subsection, we consider the tree-level EFT of the type-II seesaw model (SEFT-II), which contains one four-fermion operator
\bea
C^{}_{\ell\ell} R^{}_{\ell\ell} = \frac{(Y^{}_\Delta)_{\alpha\beta}^\ast (Y^{}_\Delta)^{}_{\gamma\delta}}{4}\left(\overline{\ell^{}_{\alpha L}} \sigma^I \epsilon\ell_{\beta L}^c\right)^\dag \left(\overline{\ell^{}_{\gamma L}} \sigma^I \epsilon \ell_{\delta L}^c\right) \;,
\eea
where the redundant operator $R^{}_{\ell\ell}$ is also equal to $2\left(\overline{\ell_{\beta i}^c}\ell_{\alpha j}\right)\left(\overline{\ell_{\gamma j}}\ell_{\delta i}^c\right)$. Then, according to Ref.~\cite{Fuentes-Martin:2022vvu}, the Fierz identities imply $R^{}_{\ell\ell} \rightarrow (1/2) Q^{}_{\ell\ell} + \text{(shifts)}$, where $Q^{}_{\ell\ell}$ is the Warsaw operator, and the ``(shifts)'' denotes the operator proportional to $\varepsilon$. One should insert the operator $R^{}_{\ell\ell} - (1/2) Q^{}_{\ell\ell}$ into one-loop diagrams and match the amplitudes to the Warsaw basis.
Since we are working in an EFT and extracting the loop divergence,
these procedures are equivalent to deriving the beta functions of the EFT operators. For this reason, we can explicitly calculate all possible on-shell amplitudes. For example, to compute the evanescent contributions to $C_{\ell q}^{\left(1,3\right)}$, one needs to consider following diagrams
\bea
\left(\begin{gathered}
	\includegraphics[width=.19\linewidth]{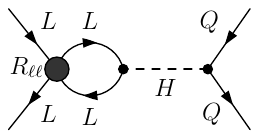}
\end{gathered}
+ \begin{gathered}
	\includegraphics[width=.19\linewidth]{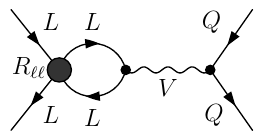}
\end{gathered}
\right)
-
\frac{1}{2}\left(
	\begin{gathered}
	\includegraphics[width=.19\linewidth]{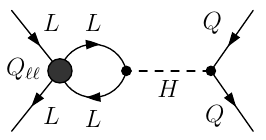}
\end{gathered}
+
\begin{gathered}
	\includegraphics[width=.19\linewidth]{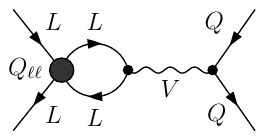}
\end{gathered}
\right) \;,
\nonumber
\eea
where $L$ and $Q$ are the lepton and quark doublets, respectively. These amplitudes give finite contributions to the process $L+L\to Q+Q$, and result in the shifts of $C_{\ell q}^{\left(1,3\right)}$ in  the SMEFT.

With a similar strategy, we insert $R^{}_{\ell\ell} - (1/2)Q^{}_{\ell\ell}$ into the one-loop diagrams and calculate the on-shell amplitudes to extract the shifts in the Wilson coefficients of the operators in the Warsaw basis. The final results are
\bea
C_{\ell q}^{\left(1\right)\alpha\beta\gamma\delta}&\rightarrow& C_{\ell q}^{\left(1\right)\alpha\beta\gamma\delta}+\frac{1}{16\pi^2}\frac{g_1^2}{18}\ \frac{\left(Y^{}_\Delta  Y_\Delta^\ast\right)^{}_{\alpha\beta}}{2}\delta^{\gamma\delta} \;, \\
C_{\ell q}^{\left(3\right)\alpha\beta\gamma\delta}&\rightarrow& C_{\ell q}^{\left(3\right)\alpha\beta\gamma\delta}-\frac{1}{16\pi^2}\frac{g_2^2}{6}\frac{\left(Y^{}_\Delta Y_\Delta^\ast\right)^{}_{\alpha\beta}}{2}\delta^{\gamma\delta} \;,\\
C_{\ell u}^{\alpha\beta\gamma\delta}&\rightarrow& C_{\ell u}^{\alpha\beta\gamma\delta}+\frac{1}{16\pi^2}\frac{2g_1^2}{9}\ \frac{\left(Y^{}_\Delta Y_\Delta^\ast\right)_{\alpha\beta}}{2}\delta^{\gamma\delta}\;,\\
C_{\ell d}^{\alpha\beta\gamma\delta}&\rightarrow& C_{\ell d}^{\alpha\beta\gamma\delta}-\frac{1}{16\pi^2}\frac{g_1^2}{9}\frac{\left(Y^{}_\Delta Y_\Delta^\ast\right)^{}_{\alpha\beta}}{2}\delta^{\gamma\delta} \;,\\
C_{H\ell}^{\left(1\right)\alpha\beta}&\rightarrow& C_{H\ell}^{\left(1\right)\alpha\beta}+\frac{1}{16\pi^2}\left[\frac{g_1^2}{6}\ \frac{\left(Y^{}_\Delta Y_\Delta^\ast\right)_{\alpha\beta}}{2}+\frac{1}{4}\left(Y^{}_\Delta Y_l^\ast Y_l^{\rm T} Y_\Delta^\ast\right)_{\alpha\beta}\right] \;,\\
C_{H\ell}^{\left(3\right)\alpha\beta}&\rightarrow& C_{H\ell}^{\left(3\right)\alpha\beta}+\frac{1}{16\pi^2}\left[-\frac{g_2^2}{6}\ \frac{\left(Y^{}_\Delta Y_\Delta^\ast\right)^{}_{\alpha\beta}}{2}+\frac{1}{4}\left(Y^{}_\Delta Y_l^\ast Y_l^{\rm T} Y_\Delta^\ast\right)^{}_{\alpha\beta}\right] \;,\\
C_{\ell e}^{\alpha\beta\gamma\delta}&\rightarrow& C_{\ell e}^{\alpha\beta\gamma\delta}+\frac{1}{16\pi^2} \left[\left(Y^{}_\Delta Y_l^\ast\right)^{}_{\alpha\gamma} \left(Y_l^{\rm T} Y_\Delta^\ast\right)^{}_{\delta\beta} - \frac{g_1^2}{3}\frac{\left(Y^{}_\Delta Y_\Delta^\ast\right)^{}_{\alpha\beta}}{2}\delta^{\gamma\delta} \right] \;,\\
C_{\ell\ell}^{\alpha\beta\gamma\delta}&\rightarrow& C_{\ell\ell}^{\alpha\beta\gamma\delta}+\frac{1}{16\pi^2}\left[\frac{3}{4}\left(g_1^2+g_2^2\right) \frac{(Y^{}_\Delta)_{\beta\delta}^\ast (Y^{}_\Delta)^{}_{\alpha\gamma}}{2}+\frac{1}{6}\left(g_2^2-g_1^2\right)\frac{\left(Y_\Delta Y_\Delta^\ast\right)_{\gamma\delta}\delta^{\alpha\beta}}{2}\right. \nonumber \\
&&\left.-\frac{g_2^2}{3}\frac{\left(Y_\Delta Y_\Delta^\ast\right)_{\gamma\beta}}{2}\delta^{\alpha\delta} \right] \;,
\eea
which can also be obtained by using the formulas in Ref.~\cite{Fuentes-Martin:2022vvu}. These new contributions have not been included in previous works on the one-loop matching of the type-II seesaw model~\cite{Li:2022ipc,Du:2022vso}. But they are necessary for the SEFT-II to reproduce the results of the full theory.

\section{Conclusions}
\label{sec:conclude}
In this work, we put forward the on-shell-amplitude approach to directly match the UV full theory at the one-loop level to the EFT operators in the physical basis. The advantage of this approach is that the Green's basis with redundant operators in the EFT is no longer necessary and thus the EOM redundance is absent. The basic strategy is to calculate all possible diagrams, impose the on-shell conditions, and match the amplitudes to the operators in the physical basis. We prove that this approach is valid and actually equivalent to the traditional one. But it is practically more efficient. One important observation is that apart from the 1LPI diagrams we have to consider also the reducible diagrams, in which the existence of the light propagator indicates the non-locality of the process. When matching the resultant on-shell amplitudes to the local operators in the EFT, we should drop all terms having momentum in the denominators. Furthermore, the approach has been implemented to carry out the one-loop matching of a toy UV model onto the EFT, to derive the beta functions in the SEFT-I, and to examine the impact of the evanescent operator on one-loop matching in the SEFT-II.

The main challenge for this approach comes from the identification of the momenta in the denominators of the amplitudes. The simplest 2-to-2 amplitude can always be expressed as a function of the Mandelstam variables $\{s, t, u\}$. Therefore, in this simple case, one will be able to obtain the matching coefficients by dropping the terms with $s$, $t$ or $u$ in the denominator. However, when the process involves more than six external legs, it is hard to find a unique set of variables, similar to the Mandelstam variables, to express the amplitudes. Consequently, it is difficult to effectively identify whether the momentum invariant in the denominator can be offset by that in the numerator. Ref.~\cite{Aebischer:2023irs} suggests using a numerical method to solve this issue, but the analytical solution is currently unavailable.

It is worthwhile to stress that apart from the one-loop matching of the UV theory, the on-shell-amplitude approach can also be applied to the EFT, such as calculating the anomalous dimensions of the operators and extracting the contribution of the evanescent operators. The approach presented in this work definitely provides a deeper insight into the EFT construction and will find more interesting applications in the phenomenological studies of the SMEFT.

\section*{Acknowledgements}
One of the authors (X.L.) would like to thank Mikael Chala for helpful discussions. This work was supported in part by the National Natural Science Foundation of China under grant No.~11835013.


\bibliography{refs}

\providecommand{\href}[2]{#2}\begingroup\raggedright\begin{thebibliography}{10}

\bibitem{Buchmuller:1985jz}
W.~Buchmuller and D.~Wyler, \emph{{Effective Lagrangian Analysis of New
  Interactions and Flavor Conservation}},
  \href{https://doi.org/10.1016/0550-3213(86)90262-2}{\emph{Nucl. Phys. B}
  {\bfseries 268} (1986) 621}.

\bibitem{Isidori:2023pyp}
G.~Isidori, F.~Wilsch and D.~Wyler, \emph{{The Standard Model effective field
  theory at work}},  \href{https://arxiv.org/abs/2303.16922}{{\ttfamily
  2303.16922}}.

\bibitem{Ellis:2018gqa}
J.~Ellis, C.W.~Murphy, V.~Sanz and T.~You, \emph{{Updated Global SMEFT Fit to
  Higgs, Diboson and Electroweak Data}},
  \href{https://doi.org/10.1007/JHEP06(2018)146}{\emph{JHEP} {\bfseries 06}
  (2018) 146} [\href{https://arxiv.org/abs/1803.03252}{{\ttfamily
  1803.03252}}].

\bibitem{Ethier:2021bye}
{\scshape SMEFiT} collaboration, \emph{{Combined SMEFT interpretation of Higgs,
  diboson, and top quark data from the LHC}},
  \href{https://doi.org/10.1007/JHEP11(2021)089}{\emph{JHEP} {\bfseries 11}
  (2021) 089} [\href{https://arxiv.org/abs/2105.00006}{{\ttfamily
  2105.00006}}].

\bibitem{deBlas:2022ofj}
J.~de~Blas, Y.~Du, C.~Grojean, J.~Gu, V.~Miralles, M.E.~Peskin et~al.,
  \emph{{Global SMEFT Fits at Future Colliders}},  in \emph{{Snowmass 2021}},
  6, 2022 [\href{https://arxiv.org/abs/2206.08326}{{\ttfamily 2206.08326}}].

\bibitem{Babu:1993qv}
K.S.~Babu, C.N.~Leung and J.T.~Pantaleone, \emph{{Renormalization of the
  neutrino mass operator}},
  \href{https://doi.org/10.1016/0370-2693(93)90801-N}{\emph{Phys. Lett. B}
  {\bfseries 319} (1993) 191}
  [\href{https://arxiv.org/abs/hep-ph/9309223}{{\ttfamily hep-ph/9309223}}].

\bibitem{Chankowski:1993tx}
P.H.~Chankowski and Z.~Pluciennik, \emph{{Renormalization group equations for
  seesaw neutrino masses}},
  \href{https://doi.org/10.1016/0370-2693(93)90330-K}{\emph{Phys. Lett. B}
  {\bfseries 316} (1993) 312}
  [\href{https://arxiv.org/abs/hep-ph/9306333}{{\ttfamily hep-ph/9306333}}].

\bibitem{Antusch:2001ck}
S.~Antusch, M.~Drees, J.~Kersten, M.~Lindner and M.~Ratz, \emph{{Neutrino mass
  operator renormalization revisited}},
  \href{https://doi.org/10.1016/S0370-2693(01)01127-3}{\emph{Phys. Lett. B}
  {\bfseries 519} (2001) 238}
  [\href{https://arxiv.org/abs/hep-ph/0108005}{{\ttfamily hep-ph/0108005}}].

\bibitem{Broncano:2004tz}
A.~Broncano, M.B.~Gavela and E.E.~Jenkins, \emph{{Renormalization of lepton
  mixing for Majorana neutrinos}},
  \href{https://doi.org/10.1016/j.nuclphysb.2004.11.001}{\emph{Nucl. Phys. B}
  {\bfseries 705} (2005) 269}
  [\href{https://arxiv.org/abs/hep-ph/0406019}{{\ttfamily hep-ph/0406019}}].

\bibitem{Jenkins:2013zja}
E.E.~Jenkins, A.V.~Manohar and M.~Trott, \emph{{Renormalization Group Evolution
  of the Standard Model Dimension Six Operators I: Formalism and lambda
  Dependence}}, \href{https://doi.org/10.1007/JHEP10(2013)087}{\emph{JHEP}
  {\bfseries 10} (2013) 087} [\href{https://arxiv.org/abs/1308.2627}{{\ttfamily
  1308.2627}}].

\bibitem{Jenkins:2013wua}
E.E.~Jenkins, A.V.~Manohar and M.~Trott, \emph{{Renormalization Group Evolution
  of the Standard Model Dimension Six Operators II: Yukawa Dependence}},
  \href{https://doi.org/10.1007/JHEP01(2014)035}{\emph{JHEP} {\bfseries 01}
  (2014) 035} [\href{https://arxiv.org/abs/1310.4838}{{\ttfamily 1310.4838}}].

\bibitem{Alonso:2013hga}
R.~Alonso, E.E.~Jenkins, A.V.~Manohar and M.~Trott, \emph{{Renormalization
  Group Evolution of the Standard Model Dimension Six Operators III: Gauge
  Coupling Dependence and Phenomenology}},
  \href{https://doi.org/10.1007/JHEP04(2014)159}{\emph{JHEP} {\bfseries 04}
  (2014) 159} [\href{https://arxiv.org/abs/1312.2014}{{\ttfamily 1312.2014}}].

\bibitem{Alonso:2014zka}
R.~Alonso, H.-M.~Chang, E.E.~Jenkins, A.V.~Manohar and B.~Shotwell,
  \emph{{Renormalization group evolution of dimension-six baryon number
  violating operators}},
  \href{https://doi.org/10.1016/j.physletb.2014.05.065}{\emph{Phys. Lett. B}
  {\bfseries 734} (2014) 302}
  [\href{https://arxiv.org/abs/1405.0486}{{\ttfamily 1405.0486}}].

\bibitem{Liao:2016hru}
Y.~Liao and X.-D.~Ma, \emph{{Renormalization Group Evolution of Dimension-seven
  Baryon- and Lepton-number-violating Operators}},
  \href{https://doi.org/10.1007/JHEP11(2016)043}{\emph{JHEP} {\bfseries 11}
  (2016) 043} [\href{https://arxiv.org/abs/1607.07309}{{\ttfamily
  1607.07309}}].

\bibitem{Liao:2019tep}
Y.~Liao and X.-D.~Ma, \emph{{Renormalization Group Evolution of Dimension-seven
  Operators in Standard Model Effective Field Theory and Relevant
  Phenomenology}}, \href{https://doi.org/10.1007/JHEP03(2019)179}{\emph{JHEP}
  {\bfseries 03} (2019) 179}
  [\href{https://arxiv.org/abs/1901.10302}{{\ttfamily 1901.10302}}].

\bibitem{Chala:2021juk}
M.~Chala and A.~Titov, \emph{{Neutrino masses in the Standard Model effective
  field theory}},
  \href{https://doi.org/10.1103/PhysRevD.104.035002}{\emph{Phys. Rev. D}
  {\bfseries 104} (2021) 035002}
  [\href{https://arxiv.org/abs/2104.08248}{{\ttfamily 2104.08248}}].

\bibitem{Chala:2021pll}
M.~Chala, G.~Guedes, M.~Ramos and J.~Santiago, \emph{{Towards the
  renormalisation of the Standard Model effective field theory to dimension
  eight: Bosonic interactions I}},
  \href{https://doi.org/10.21468/SciPostPhys.11.3.065}{\emph{SciPost Phys.}
  {\bfseries 11} (2021) 065}
  [\href{https://arxiv.org/abs/2106.05291}{{\ttfamily 2106.05291}}].

\bibitem{DasBakshi:2022mwk}
S.~Das~Bakshi, M.~Chala, A.~D\'\i{}az-Carmona and G.~Guedes, \emph{{Towards the
  renormalisation of the Standard Model effective field theory to dimension
  eight: bosonic interactions II}},
  \href{https://doi.org/10.1140/epjp/s13360-022-03194-5}{\emph{Eur. Phys. J.
  Plus} {\bfseries 137} (2022) 973}
  [\href{https://arxiv.org/abs/2205.03301}{{\ttfamily 2205.03301}}].

\bibitem{DasBakshi:2023htx}
S.~Das~Bakshi and A.~D\'\i{}az-Carmona, \emph{{Renormalisation of SMEFT bosonic
  interactions up to dimension eight by LNV operators}},
  \href{https://doi.org/10.1007/JHEP06(2023)123}{\emph{JHEP} {\bfseries 06}
  (2023) 123} [\href{https://arxiv.org/abs/2301.07151}{{\ttfamily
  2301.07151}}].

\bibitem{Weinberg:1979sa}
S.~Weinberg, \emph{{Baryon and Lepton Nonconserving Processes}},
  \href{https://doi.org/10.1103/PhysRevLett.43.1566}{\emph{Phys. Rev. Lett.}
  {\bfseries 43} (1979) 1566}.

\bibitem{Grzadkowski:2010es}
B.~Grzadkowski, M.~Iskrzynski, M.~Misiak and J.~Rosiek, \emph{{Dimension-Six
  Terms in the Standard Model Lagrangian}},
  \href{https://doi.org/10.1007/JHEP10(2010)085}{\emph{JHEP} {\bfseries 10}
  (2010) 085} [\href{https://arxiv.org/abs/1008.4884}{{\ttfamily 1008.4884}}].

\bibitem{Lehman:2014jma}
L.~Lehman, \emph{{Extending the Standard Model Effective Field Theory with the
  Complete Set of Dimension-7 Operators}},
  \href{https://doi.org/10.1103/PhysRevD.90.125023}{\emph{Phys. Rev. D}
  {\bfseries 90} (2014) 125023}
  [\href{https://arxiv.org/abs/1410.4193}{{\ttfamily 1410.4193}}].

\bibitem{Li:2020gnx}
H.-L.~Li, Z.~Ren, J.~Shu, M.-L.~Xiao, J.-H.~Yu and Y.-H.~Zheng, \emph{{Complete
  set of dimension-eight operators in the standard model effective field
  theory}}, \href{https://doi.org/10.1103/PhysRevD.104.015026}{\emph{Phys. Rev.
  D} {\bfseries 104} (2021) 015026}
  [\href{https://arxiv.org/abs/2005.00008}{{\ttfamily 2005.00008}}].

\bibitem{Murphy:2020rsh}
C.W.~Murphy, \emph{{Dimension-8 operators in the Standard Model Eective Field
  Theory}}, \href{https://doi.org/10.1007/JHEP10(2020)174}{\emph{JHEP}
  {\bfseries 10} (2020) 174}
  [\href{https://arxiv.org/abs/2005.00059}{{\ttfamily 2005.00059}}].

\bibitem{Liao:2020jmn}
Y.~Liao and X.-D.~Ma, \emph{{An explicit construction of the dimension-9
  operator basis in the standard model effective field theory}},
  \href{https://doi.org/10.1007/JHEP11(2020)152}{\emph{JHEP} {\bfseries 11}
  (2020) 152} [\href{https://arxiv.org/abs/2007.08125}{{\ttfamily
  2007.08125}}].

\bibitem{Li:2020xlh}
H.-L.~Li, Z.~Ren, M.-L.~Xiao, J.-H.~Yu and Y.-H.~Zheng, \emph{{Complete set of
  dimension-nine operators in the standard model effective field theory}},
  \href{https://doi.org/10.1103/PhysRevD.104.015025}{\emph{Phys. Rev. D}
  {\bfseries 104} (2021) 015025}
  [\href{https://arxiv.org/abs/2007.07899}{{\ttfamily 2007.07899}}].

\bibitem{Harlander:2023psl}
R.V.~Harlander, T.~Kempkens and M.C.~Schaaf, \emph{{The Standard Model
  Effective Field Theory up to Mass Dimension 12}},
  \href{https://arxiv.org/abs/2305.06832}{{\ttfamily 2305.06832}}.

\bibitem{Jiang:2018pbd}
M.~Jiang, N.~Craig, Y.-Y.~Li and D.~Sutherland, \emph{{Complete one-loop
  matching for a singlet scalar in the Standard Model EFT}},
  \href{https://doi.org/10.1007/JHEP02(2019)031}{\emph{JHEP} {\bfseries 02}
  (2019) 031} [\href{https://arxiv.org/abs/1811.08878}{{\ttfamily
  1811.08878}}].

\bibitem{Gherardi:2020det}
V.~Gherardi, D.~Marzocca and E.~Venturini, \emph{{Matching scalar leptoquarks
  to the SMEFT at one loop}},
  \href{https://doi.org/10.1007/JHEP07(2020)225}{\emph{JHEP} {\bfseries 07}
  (2020) 225} [\href{https://arxiv.org/abs/2003.12525}{{\ttfamily
  2003.12525}}].

\bibitem{Chala:2021cgt}
M.~Chala, A.~D\'\i{}az-Carmona and G.~Guedes, \emph{{A Green\textquoteright{}s
  basis for the bosonic SMEFT to dimension 8}},
  \href{https://doi.org/10.1007/JHEP05(2022)138}{\emph{JHEP} {\bfseries 05}
  (2022) 138} [\href{https://arxiv.org/abs/2112.12724}{{\ttfamily
  2112.12724}}].

\bibitem{Ren:2022tvi}
Z.~Ren and J.-H.~Yu, \emph{{A Complete Set of the Dimension-8 Green's Basis
  Operators in the Standard Model Effective Field Theory}},
  \href{https://arxiv.org/abs/2211.01420}{{\ttfamily 2211.01420}}.

\bibitem{Zhang:2023kvw}
D.~Zhang, \emph{{Renormalization Group Equations for the SMEFT Operators up to
  Dimension Seven}},  \href{https://arxiv.org/abs/2306.03008}{{\ttfamily
  2306.03008}}.

\bibitem{Chala:2020vqp}
M.~Chala and A.~Titov, \emph{{One-loop matching in the SMEFT extended with a
  sterile neutrino}},
  \href{https://doi.org/10.1007/JHEP05(2020)139}{\emph{JHEP} {\bfseries 05}
  (2020) 139} [\href{https://arxiv.org/abs/2001.07732}{{\ttfamily
  2001.07732}}].

\bibitem{Haisch:2020ahr}
U.~Haisch, M.~Ruhdorfer, E.~Salvioni, E.~Venturini and A.~Weiler,
  \emph{{Singlet night in Feynman-ville: one-loop matching of a real scalar}},
  \href{https://doi.org/10.1007/JHEP04(2020)164}{\emph{JHEP} {\bfseries 04}
  (2020) 164} [\href{https://arxiv.org/abs/2003.05936}{{\ttfamily
  2003.05936}}].

\bibitem{Dittmaier:2021fls}
S.~Dittmaier, S.~Schuhmacher and M.~Stahlhofen, \emph{{Integrating out heavy
  fields in the path integral using the background-field method: general
  formalism}},
  \href{https://doi.org/10.1140/epjc/s10052-021-09587-7}{\emph{Eur. Phys. J. C}
  {\bfseries 81} (2021) 826}
  [\href{https://arxiv.org/abs/2102.12020}{{\ttfamily 2102.12020}}].

\bibitem{Zhang:2021tsq}
D.~Zhang and S.~Zhou, \emph{{Radiative decays of charged leptons in the seesaw
  effective field theory with one-loop matching}},
  \href{https://doi.org/10.1016/j.physletb.2021.136463}{\emph{Phys. Lett. B}
  {\bfseries 819} (2021) 136463}
  [\href{https://arxiv.org/abs/2102.04954}{{\ttfamily 2102.04954}}].

\bibitem{Zhang:2021jdf}
D.~Zhang and S.~Zhou, \emph{{Complete one-loop matching of the type-I seesaw
  model onto the Standard Model effective field theory}},
  \href{https://doi.org/10.1007/JHEP09(2021)163}{\emph{JHEP} {\bfseries 09}
  (2021) 163} [\href{https://arxiv.org/abs/2107.12133}{{\ttfamily
  2107.12133}}].

\bibitem{Coy:2021hyr}
R.~Coy and M.~Frigerio, \emph{{Effective comparison of neutrino-mass models}},
  \href{https://doi.org/10.1103/PhysRevD.105.115041}{\emph{Phys. Rev. D}
  {\bfseries 105} (2022) 115041}
  [\href{https://arxiv.org/abs/2110.09126}{{\ttfamily 2110.09126}}].

\bibitem{Ohlsson:2022hfl}
T.~Ohlsson and M.~Pernow, \emph{{One-loop matching conditions in neutrino
  effective theory}},
  \href{https://doi.org/10.1016/j.nuclphysb.2022.115729}{\emph{Nucl. Phys. B}
  {\bfseries 978} (2022) 115729}
  [\href{https://arxiv.org/abs/2201.00840}{{\ttfamily 2201.00840}}].

\bibitem{Li:2022ipc}
X.~Li, D.~Zhang and S.~Zhou, \emph{{One-loop matching of the type-II seesaw
  model onto the Standard Model effective field theory}},
  \href{https://doi.org/10.1007/JHEP04(2022)038}{\emph{JHEP} {\bfseries 04}
  (2022) 038} [\href{https://arxiv.org/abs/2201.05082}{{\ttfamily
  2201.05082}}].

\bibitem{Du:2022vso}
Y.~Du, X.-X.~Li and J.-H.~Yu, \emph{{Neutrino seesaw models at one-loop
  matching: Discrimination by effective operators}},
  \href{https://arxiv.org/abs/2201.04646}{{\ttfamily 2201.04646}}.

\bibitem{Zhang:2022osj}
D.~Zhang, \emph{{Complete one-loop structure of the type-(I+II) seesaw
  effective field theory}},
  \href{https://doi.org/10.1007/JHEP03(2023)217}{\emph{JHEP} {\bfseries 03}
  (2023) 217} [\href{https://arxiv.org/abs/2208.07869}{{\ttfamily
  2208.07869}}].

\bibitem{Liao:2022cwh}
Y.~Liao and X.-D.~Ma, \emph{{One-loop matching of scotogenic model onto
  standard model effective field theory up to dimension 7}},
  \href{https://doi.org/10.1007/JHEP12(2022)053}{\emph{JHEP} {\bfseries 12}
  (2022) 053} [\href{https://arxiv.org/abs/2210.04270}{{\ttfamily
  2210.04270}}].

\bibitem{Fuentes-Martin:2020udw}
J.~Fuentes-Martin, M.~K\"onig, J.~Pag\`es, A.E.~Thomsen and F.~Wilsch,
  \emph{{SuperTracer: A Calculator of Functional Supertraces for One-Loop EFT
  Matching}}, \href{https://doi.org/10.1007/JHEP04(2021)281}{\emph{JHEP}
  {\bfseries 04} (2021) 281}
  [\href{https://arxiv.org/abs/2012.08506}{{\ttfamily 2012.08506}}].

\bibitem{Carmona:2021xtq}
A.~Carmona, A.~Lazopoulos, P.~Olgoso and J.~Santiago, \emph{{Matchmakereft:
  automated tree-level and one-loop matching}},
  \href{https://doi.org/10.21468/SciPostPhys.12.6.198}{\emph{SciPost Phys.}
  {\bfseries 12} (2022) 198}
  [\href{https://arxiv.org/abs/2112.10787}{{\ttfamily 2112.10787}}].

\bibitem{Fuentes-Martin:2022jrf}
J.~Fuentes-Mart\'\i{}n, M.~K\"onig, J.~Pag\`es, A.E.~Thomsen and F.~Wilsch,
  \emph{{A Proof of Concept for Matchete: An Automated Tool for Matching
  Effective Theories}},  \href{https://arxiv.org/abs/2212.04510}{{\ttfamily
  2212.04510}}.

\bibitem{Aebischer:2023irs}
J.~Aebischer et~al., \emph{{Computing Tools for Effective Field Theories}},  7,
  2023 [\href{https://arxiv.org/abs/2307.08745}{{\ttfamily 2307.08745}}].

\bibitem{Zhang:2016pja}
Z.~Zhang, \emph{{Covariant diagrams for one-loop matching}},
  \href{https://doi.org/10.1007/JHEP05(2017)152}{\emph{JHEP} {\bfseries 05}
  (2017) 152} [\href{https://arxiv.org/abs/1610.00710}{{\ttfamily
  1610.00710}}].

\bibitem{Beneke:1997zp}
M.~Beneke and V.A.~Smirnov, \emph{{Asymptotic expansion of Feynman integrals
  near threshold}},
  \href{https://doi.org/10.1016/S0550-3213(98)00138-2}{\emph{Nucl. Phys. B}
  {\bfseries 522} (1998) 321}
  [\href{https://arxiv.org/abs/hep-ph/9711391}{{\ttfamily hep-ph/9711391}}].

\bibitem{Jantzen:2011nz}
B.~Jantzen, \emph{{Foundation and generalization of the expansion by regions}},
  \href{https://doi.org/10.1007/JHEP12(2011)076}{\emph{JHEP} {\bfseries 12}
  (2011) 076} [\href{https://arxiv.org/abs/1111.2589}{{\ttfamily 1111.2589}}].

\bibitem{Cohen:2022uuw}
T.~Cohen, N.~Craig, X.~Lu and D.~Sutherland, \emph{{On-Shell Covariance of
  Quantum Field Theory Amplitudes}},
  \href{https://doi.org/10.1103/PhysRevLett.130.041603}{\emph{Phys. Rev. Lett.}
  {\bfseries 130} (2023) 041603}
  [\href{https://arxiv.org/abs/2202.06965}{{\ttfamily 2202.06965}}].

\bibitem{Arzt:1993gz}
C.~Arzt, \emph{{Reduced effective Lagrangians}},
  \href{https://doi.org/10.1016/0370-2693(94)01419-D}{\emph{Phys. Lett. B}
  {\bfseries 342} (1995) 189}
  [\href{https://arxiv.org/abs/hep-ph/9304230}{{\ttfamily hep-ph/9304230}}].

\bibitem{Henning:2016lyp}
B.~Henning, X.~Lu and H.~Murayama, \emph{{One-loop Matching and Running with
  Covariant Derivative Expansion}},
  \href{https://doi.org/10.1007/JHEP01(2018)123}{\emph{JHEP} {\bfseries 01}
  (2018) 123} [\href{https://arxiv.org/abs/1604.01019}{{\ttfamily
  1604.01019}}].

\bibitem{Fuentes-Martin:2016uol}
J.~Fuentes-Martin, J.~Portoles and P.~Ruiz-Femenia, \emph{{Integrating out
  heavy particles with functional methods: a simplified framework}},
  \href{https://doi.org/10.1007/JHEP09(2016)156}{\emph{JHEP} {\bfseries 09}
  (2016) 156} [\href{https://arxiv.org/abs/1607.02142}{{\ttfamily
  1607.02142}}].

\bibitem{DeAngelis:2023bmd}
S.~De~Angelis and G.~Durieux, \emph{{EFT matching from analyticity and
  unitarity}},  \href{https://arxiv.org/abs/2308.00035}{{\ttfamily
  2308.00035}}.

\bibitem{Li:2022aby}
X.~Li, \emph{{Positivity bounds at one-loop level: the Higgs sector}},
  \href{https://doi.org/10.1007/JHEP05(2023)230}{\emph{JHEP} {\bfseries 05}
  (2023) 230} [\href{https://arxiv.org/abs/2212.12227}{{\ttfamily
  2212.12227}}].

\bibitem{Chala:2021wpj}
M.~Chala and J.~Santiago, \emph{{Positivity bounds in the standard model
  effective field theory beyond tree level}},
  \href{https://doi.org/10.1103/PhysRevD.105.L111901}{\emph{Phys. Rev. D}
  {\bfseries 105} (2022) L111901}
  [\href{https://arxiv.org/abs/2110.01624}{{\ttfamily 2110.01624}}].

\bibitem{Davidson:2018zuo}
S.~Davidson, M.~Gorbahn and M.~Leak, \emph{{Majorana neutrino masses in the
  renormalization group equations for lepton flavor violation}},
  \href{https://doi.org/10.1103/PhysRevD.98.095014}{\emph{Phys. Rev. D}
  {\bfseries 98} (2018) 095014}
  [\href{https://arxiv.org/abs/1807.04283}{{\ttfamily 1807.04283}}].

\bibitem{Wang:2023bdw}
Y.~Wang, D.~Zhang and S.~Zhou, \emph{{Complete one-loop renormalization-group
  equations in the seesaw effective field theories}},
  \href{https://doi.org/10.1007/JHEP05(2023)044}{\emph{JHEP} {\bfseries 05}
  (2023) 044} [\href{https://arxiv.org/abs/2302.08140}{{\ttfamily
  2302.08140}}].

\bibitem{Fuentes-Martin:2022vvu}
J.~Fuentes-Mart\'\i{}n, M.~K\"onig, J.~Pag\`es, A.E.~Thomsen and F.~Wilsch,
  \emph{{Evanescent operators in one-loop matching computations}},
  \href{https://doi.org/10.1007/JHEP02(2023)031}{\emph{JHEP} {\bfseries 02}
  (2023) 031} [\href{https://arxiv.org/abs/2211.09144}{{\ttfamily
  2211.09144}}].

\end{thebibliography}\endgroup
\bibliographystyle{JHEP}

\end{document}